\documentclass[12pt,preprint]{elsarticle}

\usepackage{lineno,hyperref,amssymb,graphicx,float,subcaption}
\usepackage{amsmath,bm, amsthm, mathtools}
\usepackage[ruled,vlined]{algorithm2e}
\SetKwComment{Comment}{$\triangleright$\ }{}
\usepackage{color}
\usepackage{geometry}
\usepackage{pdflscape}
\modulolinenumbers[5]
\graphicspath{{figures/}}
\theoremstyle{remark}
\newdefinition{remark}{Remark}
\newtheorem{prop}{Proposition}
\theoremstyle{definition}
\newtheorem{definition}{Definition}[section]

\newcommand{\etal}{\textit{et al}.}

\newcommand{\viscosity}{\nu}

\newcommand{\mc}{\mathcal}
\def\floor#1{\lfloor #1 \rfloor}
\newcommand{\norm}[1]{\left\lVert #1 \right\rVert}

\DeclareMathOperator*{\argmin}{arg\,min}
\def\Eqref#1{Eq.~(\ref{#1})}

\journal{Journal of Computational Physics}

\begin{document}

\begin{frontmatter}

\title{Modeling the Dynamics of PDE Systems with Physics-Constrained Deep Auto-Regressive Networks}

 \author[label1]{Nicholas Geneva}
 \ead{ngeneva@nd.edu}

 \author[label1]{Nicholas Zabaras\corref{cor1}}
 \ead{nzabaras@gmail.com}
 \ead[url]{https://cics.nd.edu/}

\address[label1]{Center for Informatics and Computational Science, University of Notre Dame, 311 Cushing Hall, Notre Dame, IN 46556, USA}
\cortext[cor1]{Corresponding author}

\begin{abstract}
    In recent years, deep learning has proven to be a viable methodology for surrogate modeling and uncertainty quantification for a vast number of physical systems.
    However, in their traditional form, such models can require a large amount of training data.
    This is of particular importance for various engineering and scientific applications where data may be extremely expensive to obtain.
    To overcome this shortcoming, physics-constrained deep learning provides a promising methodology as it only utilizes the   governing equations.
    In this work, we propose a novel auto-regressive dense encoder-decoder convolutional neural network   to solve and model non-linear dynamical systems without training data at a computational cost   that is potentially magnitudes lower than standard numerical solvers. 
    This model includes a Bayesian framework that allows for uncertainty quantification of the predicted quantities of interest at each time-step.
    We rigorously test this model on several non-linear transient partial differential equation systems including the turbulence of the Kuramoto-Sivashinsky equation, multi-shock formation and interaction with 1D Burgers' equation and 2D wave dynamics with coupled Burgers' equations.
    For each system, the predictive results and uncertainty are presented and discussed together with comparisons to the results obtained from traditional numerical analysis methods.
\end{abstract}

\begin{keyword}
Physics-Informed Machine Learning  \sep Auto-regressive Model     \sep Deep Neural Networks \sep Convolutional Encoder-Decoder \sep Uncertainty Quantification \sep Dynamics \sep Partial Differential Equations
\end{keyword}

\end{frontmatter}

%\linenumbers

% ==== Introduction ====
\section{Introduction}
\noindent
In almost all scientific domains,  simulating systems of partial differential equations (PDEs) is of great importance and research interest. 
Given that many physical phenomena including heat diffusion, fluid dynamics, and elasticity are formalized with PDEs, numerically or analytically solving these governing equations is a core foundation for a vast spectrum of scientific and engineering disciplines. 
In recent decades exponential growth in computational power has made such numerical methods for solving PDEs even more accessible. 
However, in most modern-day applications, obtaining the desired resolution or accuracy with such simulations is still computationally expensive. 
Hence, many seek to strike an ideal balance between predictive accuracy and computational efficiency.
In many situations, such as optimization or inverse problems, a large number of repeated simulations are required prioritizing the computational efficiency of the numerical simulator.
Often surrogate models are used to ease this computational burden by providing a fast approximate model that can imitate a standard numerical solver at a significantly reduced computational cost.

In recent years, machine learning and   deep learning have entered a renaissance in which groundbreaking findings have made deep learning models widely successful for a vast number of applications~\cite{lecun2015deep}.
One such application is surrogate modeling in which a deep learning model can be used as a black box method to approximate a physical system.
Among the most popular deep learning models is deep neural networks (DNNs), which have proven to be an extremely effective method for modeling a wide spectrum of physical systems such as flow through porous media~\cite{zhu2018bayesian, zhu2019physics, tripathy2018deep}, Navier-Stokes equations~\cite{yang2016data}, turbulence modeling~\cite{geneva2019quantifying}, molecular dynamics~\cite{schoberl2019predictive} and more.
Traditional DNNs are not probabilistic in nature resulting in Bayesian extensions of these models~\cite{mackay1992practical,neal2012bayesian} to quantify the underlying uncertainty in these black box algorithms.
While DNNs have been proven to be both accurate and computationally efficient for modeling and uncertainty quantification (UQ), it is commonly known that training such models may require a significant amount of   data. 
Depending on the system of interest, training data may either be sparse, extremely expensive to obtain or not available at all. 
Considering that the   underlying governing equations are known, in this work, we are particularly interested in the surrogate modeling of physical systems using physics-constrained loss functions.
Such loss functions allow a surrogate model to be trained in the \textit{absence} of data (e.g. without having to solve the equations governing the system of interest). 

The philosophy of learning ordinary or partial differential equations through constraint based loss functions is far from a new idea with related works reaching back over two decades ago~\cite{psichogios1992hybrid, meade1994numerical, meade1994solution, lagaris1998artificial}. 
These early works focused on solving initial/boundary value problems in which the solution is parameterized by a fully-connected network  which allows for a fully differentiable and closed analytic form~\cite{lagaris1998artificial}.
With the resurgence of interest in neural networks, such techniques have been rediscovered by multiple works in recent years where this core idea has been expanded upon.
As discussed in the work of Lagaris~\etal~\cite{lagaris1998artificial, lagaris2000neural} and later revisited by Raissi~\etal~\cite{raissi2019physics}, the use of fully connected networks with physics-constrained learning allows 
for a mesh free solution that can be evaluated anywhere on the domain while being trained on only a few points.
Additionally, Lagaris~\etal~\cite{lagaris2000neural} and more recently Berg and Nystr{\"o}m~\cite{berg2018unified} showed fully connected networks can be used to learn PDE solutions on even complex domains.
Recently, several investigators have examined the use of variational formulations of the governing equations as loss functions to solve various PDEs~\cite{zhu2019physics, weinan2018deep, nabian2018deep, karumuri2019simulator} which has been proven to be effective.
Sirignano~\etal~\cite{sirignano2018dgm} show that the use of a fully connected network can be used for efficiently solving PDEs of high dimensionality where traditional
discretization techniques become unfeasible.
Several have also investigated the use of fully connected networks to solve high-dimensional stochastic PDEs with good success~\cite{grohs2018proof, khoo2019solving}.
While fully connected networks could be optimized to compute a single solution of a PDE, several challenges remain in extending these ideas
to surrogate model construction.  
For example, if the initial condition, boundary conditions, material properties, etc. are changed the model must be retrained.
This means that from a computational aspect, such methods are difficult to justify if a numerical simulator can be used that is computationally less expensive than training the fully connected network.
Clearly, with decades of numerical analysis progress this issue is applicable to an overwhelmingly large amount of PDE systems.

To the authors best knowledge only the works of Zhu~\etal~\cite{zhu2019physics} and Karumuri~\etal~\cite{karumuri2019simulator} seek to build surrogate models using physics-constrained, data free learning.
In~\cite{zhu2019physics}, a deep convolutional neural network was used to formulate a surrogate model for an elliptic PDE with a stochastic, high-dimensional permeability field.
Additionally, Zhu~\etal~proposed a probabilistic framework based on a conditional flow-based generative model~\cite{dinh2016density} to quantify the potential error arising from the model itself.
It was also found that the data-less physics-constrained learning yielded a model with much better generalization capabilities than traditional data-driven learning.
Karumuri~\etal~\cite{karumuri2019simulator} used a deep fully connected ResNet~\cite{he2016deep} to build a surrogate also for elliptic PDEs with reasonable success.
Note that both of these aforementioned works have been focused entirely on PDEs which are not time-dependent in nature.

In this work, we generalize these physics-constrained deep learning surrogate models to dynamical PDEs.
The novel contributions of this paper are as follows: 
(a) We propose a deep auto-regressive dense encoder-decoder for predicting transient PDEs and the physics-constrained training algorithm;
(b) Extend this model to a Bayesian framework using the recently proposed stochastic weight averaging Gaussian algorithm to quantify both epistemic and aleatoric uncertainty; 
(c) Implement this model for a chaotic/turbulent system, a system with multiple shock wave interactions and a 2D system of coupled non-linear PDEs far surpassing the complexity of other test cases shown in past literature; 
(d) Present and discuss the accuracy of the predictions as well as the associated uncertainty for each of the previously discussed PDEs; and 
(e) Compare the computational efficiency of the proposed surrogate model against other state-of-the-art numerical methods.

This paper is organized as follows: First, in Section~\ref{sec:prob-definition}, we briefly define and discuss the problem of interest.
Section~\ref{sec:ar-denseed} discusses the auto-regressive dense encoder-decoder model, its training and use as a surrogate model.
In Section~\ref{sec:bar-denseed}, we extend this deep learning model to the Bayesian paradigm where we discuss the formulation of the posterior as well as the approximation of the predictive distribution.
Following, in Section~\ref{sec:ks}, the proposed model is implemented for the chaotic Kuramoto-Sivashinsky system and a study is presented of the turbulent statistics that the model produces.
In Section~\ref{sec:1dVisBurgers}, we also explore the use of the auto-regressive model for the prediction of shocks in the 1D Burgers' equation.
Later in Section~\ref{sec:2dVisBurgers}, we further extend this to the 2D coupled Burgers' system.
Lastly, conclusions and  discussion can be found in Section~\ref{sec:conclusion}.
All code, trained models and data used in this work is open-sourced for full reproducibility.\footnote{Code available at: \href{https://github.com/cics-nd/ar-pde-cnn}{https://github.com/cics-nd/ar-pde-cnn}.}

% ========= Problem Definition =========
\section{Problem Definition}
\noindent
\label{sec:prob-definition}
In this work, we are interested in using deep learning architectures for developing surrogate models of non-linear dynamical systems that evolve in both space and time using physics-constrained learning.
Specifically, we wish to use the governing equations to formulate loss functions for training surrogate models without the need of (output) training data.
Our goal is to develop surrogate models for a class of arbitrary transient PDE systems with an unknown, variable or stochastic initial state.
Consider a transient system of PDEs that models a physical system:
\begin{equation}
    \begin{gathered}
    \bm{u}(\bm{x},t)_{t}+F\left(\bm{x}, \bm{u}\left(\bm{x},t\right)\right) = 0, \quad  \bm{x}\in\Omega, \: t\in[0,T],\\ 
    \mathcal{B}(\bm{u}) = b\left(\bm{x},t\right), \quad \bm{x}\in\Gamma,\\
    \bm{u}(\bm{x},0) \sim p\left(\bm{u}(\bm{x},0)\right),
    \end{gathered}
    \label{eq:general-pde}
\end{equation}
where we have denoted this $n$-dimensional PDE by the temporal derivative $\bm{u}(\bm{x},t)_{t}$ and the remaining terms by $F(\cdot)$ which includes spatial derivatives and non-linear terms.
$\bm{u}(\bm{x},t)$ are the system's state variables in the domain $\Omega$ with a boundary $\Gamma$.
$\mathcal{B}$ is the boundary operator that enforces the desired boundary conditions.
Lastly, the initial state $\bm{u}(\bm{x},0)$ is a real valued random field with a probability density, $p\left(\bm{u}(\bm{x},0)\right)$, that may or may not be known.

Our goal is to expand on the work in~\cite{zhu2019physics} in which PDE solutions were represented as an optimization problem by either minimizing an energy functional or alternatively the square of the PDE residual~\cite{filippov1992variational}.
The objective in~\cite{zhu2019physics} was to predict quantities of interest for an elliptic PDE (defining Darcy's flow) in an image-to-image regression approach using a convolutional encoder-decoder architecture with an input being a property field 
(permeability) and the output being the quantities of interest (pressure and velocity).
The use of a convolutional neural network proved to have some significant benefits over the more commonly used fully-connected networks including faster convergence and better accuracy.
While successful for elliptic PDEs, the strategies in this past work  cannot directly generalize  to a dynamical system.

If one were developing a numerical algorithm to solve a dynamical system, the first step would be to discretize the time derivative, which is commonly referred to as a time-stepping or time integration method~\cite{ralston2001first}.
For time integration there are a vast number of options including standard explicit or implicit methods, Runge-Kutta methods, linear multi-step methods, implicit-explicit methods and more.
However, the goal of all these techniques is the same: evolve the system from time 
$t$ to time $t+\Delta t$.
Using this philosophy of discrete time integration, we propose building a surrogate model that performs time integration at a specified $\Delta t$ in an image-to-image regression algorithm using a convolutional encoder-decoder neural network.
Let us consider $\bm{u}^{n}$ as the solution of the PDE with $d_{0}$ state variables on a given  structured Euclidean discretization of $\Omega$ at time-step $n$.
Namely, given $\Omega$ discretized with $D_{i}$ points in the $i$-th dimension, $\bm{u}^{n}\in\mathbb{R}^{d_{0} \times D_{1}}$ for a 1D system, $\bm{u}^{n}\in\mathbb{R}^{d_{0} \times D_{1} \times D_{2}}$ for a 2D system and  $\bm{u}^{n}\in\mathbb{R}^{d_{0} \times D_{1} \times D_{2} \times D_{3}}$ for a 3D system.
Our convolutional encoder-decoder model for simulating time integration at time-step $n$ is parameterized as follows:
\begin{equation}
    \bm{u}^{n+1} = f\left(\bm{\chi}^{n+1}, \textbf{w}\right), \quad \bm{\chi}^{n+1} \equiv \left\{\bm{u}^{n}, \bm{u}^{n-1}, \ldots, \bm{u}^{n-k} \right\},
    \label{eq:time-integration}
\end{equation}
where $f$ represents the function learned by the deep learning model, $\textbf{w}$ are the learnable parameters in this convolutional neural network.
$\bm{\chi}^{n+1}$ is the model's input, for the prediction $\bm{u}^{n+1}$, consisting of the $k+1$ previous states of the system.
By this model definition, we are interested in learning the dynamics or evolution of the system invariant to the current time $t$.
The use of a convolutional neural network allows for a light-weight model that can evolve the system of interest by a discrete time-step efficiently without any matrix inversions, iterative relaxations or multi-step processes.
Similar to the convolutional model in Zhu~\etal~\cite{zhu2019physics}, this model can be used/extended for tasks such as solving PDEs, surrogate modeling and performing uncertainty quantification.

To predict a given system's response for $N$ time-steps, the convolutional neural network is executed as an auto-regressive model.
Given just a discretized initial state of the system $\bm{u}_{0}$, one can predict the system response as:
\begin{align}
\begin{split}
    \bm{u}^{1} = f\left(\bm{\chi}^{1}, \textbf{w}\right), &\quad \bm{\chi}^{1} = \left\{\bm{u}_{0}, \bm{u}_{0}, \bm{u}_{0}, \ldots, \bm{u}_{0} \right\},\\
    \bm{u}^{2} = f\left(\bm{\chi}^{2}, \textbf{w}\right), &\quad \bm{\chi}^{2} = \left\{\bm{u}^{1}, \bm{u}_{0}, \bm{u}_{0}, \ldots, \bm{u}_{0} \right\},\\
    \bm{u}^{3} = f\left(\bm{\chi}^{3}, \textbf{w}\right), &\quad \bm{\chi}^{3} = \left\{\bm{u}^{2}, \bm{u}^{1}, \bm{u}_{0}, \ldots, \bm{u}_{0} \right\},\\
    \ldots \\
    \bm{u}^{N} = f\left(\bm{\chi}^{N}, \textbf{w}\right), &\quad \bm{\chi}^{N} = \left\{\bm{u}^{N-1}, \bm{u}^{N-2}, \ldots, \bm{u}^{N-1-k} \right\},
    \label{eq:time-int}
\end{split}
\end{align}
where the model must be executed $N$ times to obtain the prediction of the system at time-step $N$.
Note, that the initial input to the model $\bm{\chi}^{1}$ is comprised of just the initial state, which is an approximation needed to ``kick-start'' the time series.
The prediction for a particular time-step can be formulated as a recursive function of the model:
\begin{equation}
    \bm{u}^{n+1} = f\left(\left\{ f\left(\bm{\chi}^{n}, \textbf{w}\right), f\left(\bm{\chi}^{n-1}, \textbf{w}\right), \ldots, f\left(\bm{\chi}^{n-k}, \textbf{w}\right) \right\}, \textbf{w}\right),
    \label{eq:recursive}
\end{equation}
where the input $\bm{\chi}^{n+1}$ is formulated in terms of the model itself, with inputs that can be described in a similar manner.
Thus only the initial state is needed for predicting a systems' response up to an arbitrary number of time-steps.
For the prediction of an entire time series $\left[\bm{u}^{N}, \bm{u}^{N-1}, \ldots \bm{u}^{1}\right]$, the model can be represented as a set of functions $\hat{f}(\bm{u}_{0}, \textbf{w})=\left\{f\left(\bm{\chi}^{N}, \textbf{w}\right), f\left(\bm{\chi}^{N-1}, \textbf{w}\right), \ldots, f\left(\bm{\chi}^{1}, \textbf{w}\right)\right\}$ in which each can be expressed recursively as a function of the initial state.

As discussed previously, we would like to formulate a methodology of physics-constrained learning such that the model can be trained with only a set of initial states $\bm{u}_{0}$.
Although the same model is used to predict every time-step of a given time series as shown in~\Eqref{eq:time-int}, the core building block for training this DNN model is learning how to predict the transition of states from $t$ to $t+\Delta t$ regardless of the reference time $t$.
For physics-constrained learning, we will pose the optimization problem for a single time-step, $n\rightarrow n+1$, as the minimization of the discrepancy between the model's prediction 
and the prediction of a discrete numerical time integration method $T_{\Delta t}$ of time-step $\Delta t$ for the governing equation of interest:
\begin{equation}
    \argmin_{\textbf{w}} \norm{f\left(\bm{\chi}^{n+1}, \textbf{w}\right) -  T_{\Delta t}\left(\bm{\mathcal{U}}^{n+1}, F_{\Delta x}(\cdot)\right)}^{2}_{2},
    \label{eq:general-loss}
\end{equation}
in which the systems states $\bm{u}^{n+1}$ are parameterized by the DNN. 
The time-integrator predicts numerically the state at the next time-step given some system states, $\bm{\mathcal{U}}^{n+1}$, and a discretized form of the additional terms of the PDE $F_{\Delta x}$.
The exact definition of $\bm{\mathcal{U}}^{n+1}$ depends on the time integration method used.
For example, for the explicit forward Euler scheme, $\bm{\mathcal{U}}^{n+1} = \left\{\bm{u}^{n}\right\}$ depends only on the previous time-step state while for the Crank-Nicolson scheme, $\bm{\mathcal{U}}^{n+1} = \left\{\bm{u}^{n+1}, \bm{u}^{n}\right\}$, 
depends on \textit{both} the model's current prediction as well as the state at the previous time-step.
As it will be discussed in greater detail in the following sections, given that $T_{\Delta t}$ is consistent with the governing equation, this minimization can be interpreted as the minimization of the residual of the discretized PDE.

\subsection{Surrogate Modeling of Dynamical Systems}
\noindent
In the context of this work, we are focused on developing a surrogate model that can efficiently predict a dynamical system's response  $\bm{y}=\left[\bm{u}^{N}, \bm{u}^{N-1}, \ldots \bm{u}^{1}\right]$ for time-steps $[1,N]$ for a given initial state realization 
$\bm{u}_{0,i} \sim p\left(\bm{u}_{0}\right)$ and a set of boundary conditions.
We pose the following definition for this surrogate model:
\theoremstyle{definition}
\begin{definition}{(Deterministic Surrogate Model)}
    For a transient PDE system with specified boundary conditions, as in~\Eqref{eq:general-pde}, and a finite set of initial conditions $\mathcal{S}=\left\{\bm{u}_{0,i}\right\}_{i=1}^{M}$, $\bm{u}_{0,i} \sim p\left(\bm{u}_{0}\right)$, train a surrogate to predict the dynamical response $\bm{y}=\hat{f}\left(\bm{u}_{0}^{*}, \textbf{w}\right)$ for any initial condition, $\bm{u}_{0}^{*} \sim p\left(\bm{u}_{0}\right)$, such that the predicted response is the solution of the governing PDEs for the respective initial state.
\end{definition}
\noindent
The true density of $p\left(\bm{u}_{0}\right)$ may not be known.
For example, $p\left(\bm{u}_{0}\right)$ may represent a set of states collected from an experiment or simulation.
When this density is not known or samples are not available, one may need to approximate it for the sake of assembling a set of initial states for training.
In the context of this work, we will pose this initial condition as a random function from which we can sample from to illustrate the applicability of our model.
As discussed in Zhu~\etal~\cite{zhu2019physics}, surrogate modeling using physics-constrained learning can be interpreted as unsupervised learning since training takes place without any   labeled training data.
Rather it is up to the model to discover the dynamics of the system.

In the majority of problems of interest, the number of (initial) training data used will only express a portion of all the inputs that can be drawn from the density function $p\left(\bm{u}_{0}\right)$.
Thus the surrogate model will only be trained to predict a part of all potential responses of the dynamical system.
To account for this limited expressibility of both the input data used for training as well as the trained model itself, we also wish to formulate a probabilistic surrogate than can produce distributions over possible solutions, rather than a single point estimate.
Hence, we pose the following definition for the probabilistic extension of this dynamical surrogate model:
\theoremstyle{definition}
\begin{definition}{(Probabilistic Surrogate Model)}
    For a transient PDE system with specified boundary conditions, such as~\Eqref{eq:general-pde}, and a finite set of initial conditions $\mathcal{S}=\left\{\bm{u}_{0,i}\right\}_{i=1}^{M}$, $\bm{u}_{0,i} \sim p\left(\bm{u}_{0}\right)$, train a surrogate to predict the dynamical response density, $p\left(\bm{y}| \bm{u}_{0}^{*}, \mathcal{S}\right)$, such that the samples drawn from the predictive distribution satisfy the governing PDEs for the respective initial state.
\end{definition}

% ========= Auto-regressive Dense Encoder-Decoder =========
\section{Auto-regressive Dense Encoder-Decoder}
\label{sec:ar-denseed}
\begin{figure}[H]
    \centering
    \includegraphics[trim={0 0.25cm 0 0},clip, width=0.9\textwidth]{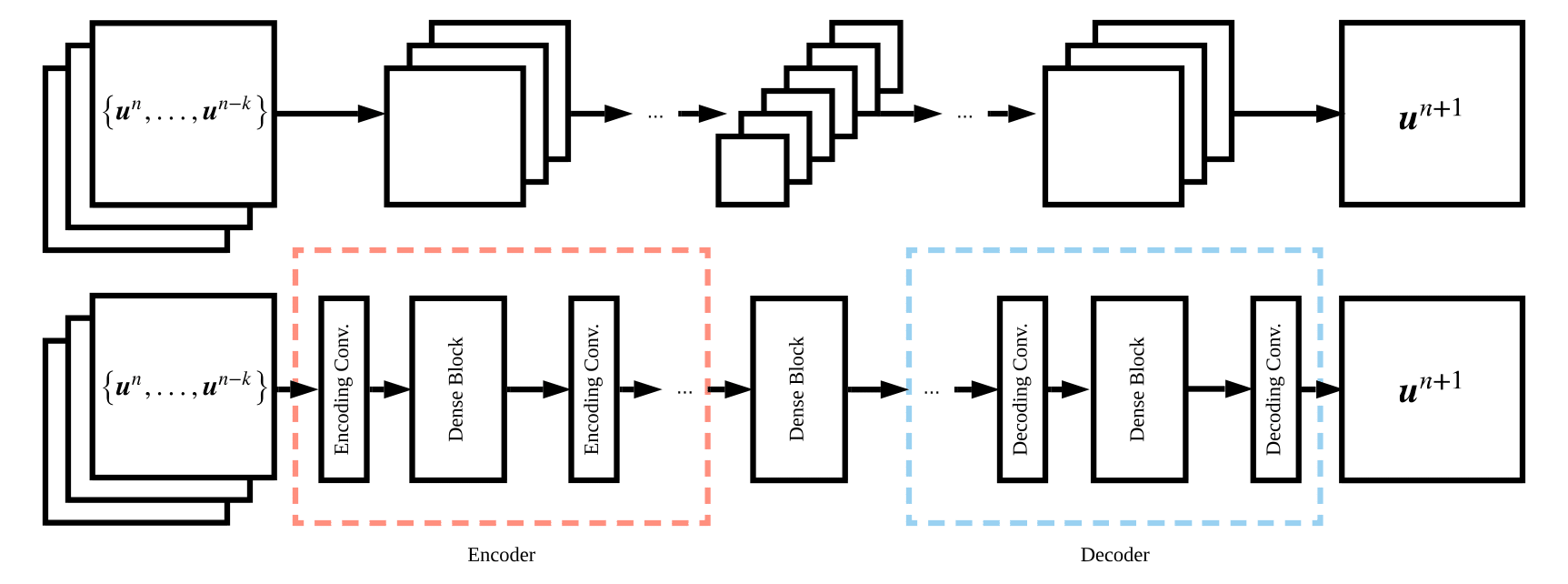}
    \caption{Schematic of the auto-regressive dense encoder-decoder. The top shows the dimensionality of the data in the network, the bottom shows the model architecture.
    Using encoding convolutions lowers the dimensionality of the input feature map, while decoding convolutions increase the dimensionality.
    The encoding-decoding process is interleaved with dense blocks that contain multiple densely connected layers in which the dimensionality of the feature maps is held constant.
    Additional details on these components can be found in~\ref{app:ks} and the work of Zhu and Zabaras~\cite{zhu2018bayesian}.}
    \label{fig:ar-denseED}
\end{figure}

\begin{figure}[H]
    \centering
    \includegraphics[trim={0 0.25cm 0 0},clip, width=0.7\textwidth]{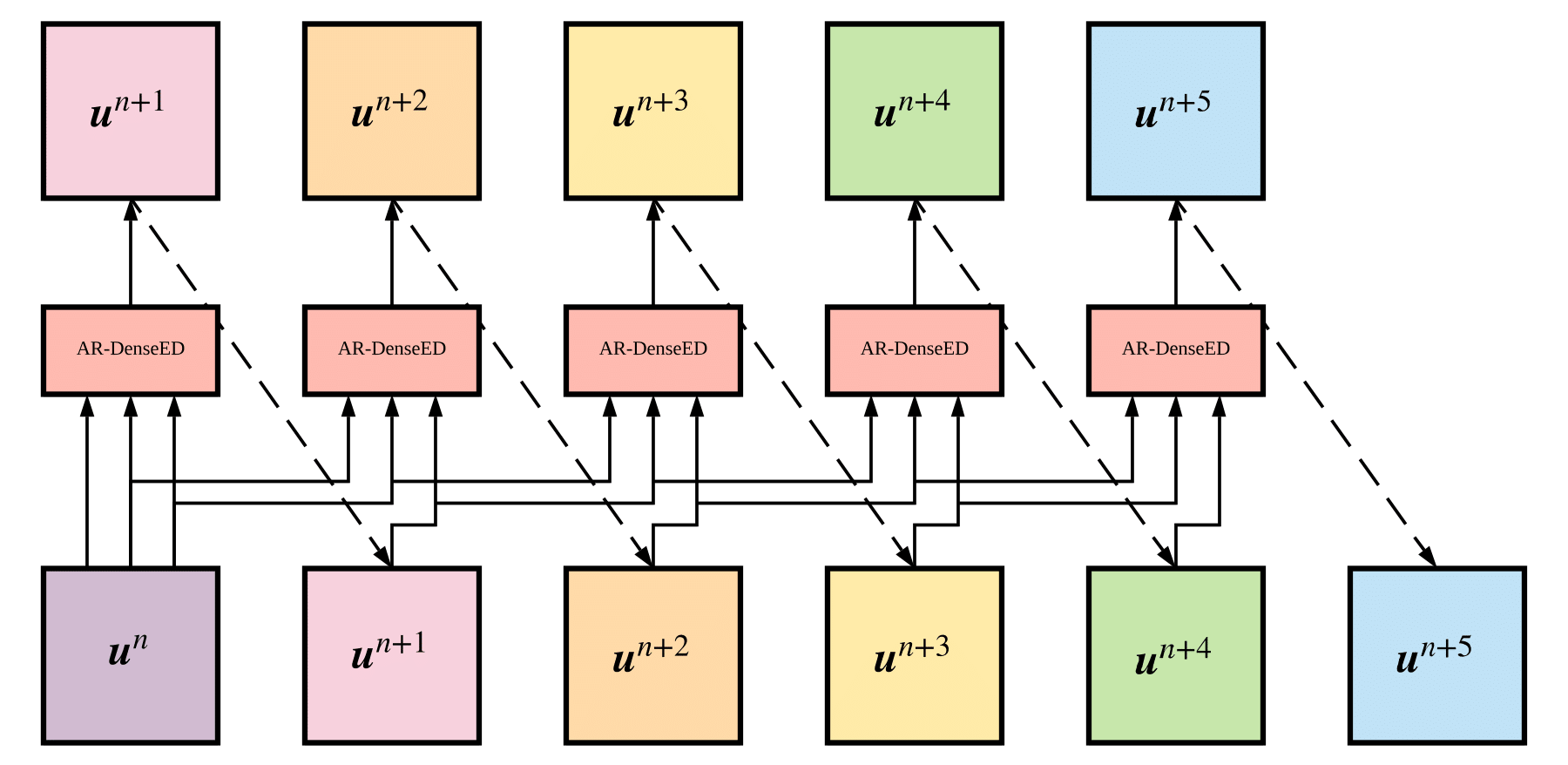}
    \caption{AR-DenseED prediction, outlined in Algorithm~\ref{algo:testing}, of the states at five uniformly spaced time-steps using the states at three previous time-steps as inputs.
    During prediction, the model used is identical for each time-step.
    At the beginning of the time sequence, all three inputs are the initial state $\bm{u}^{n}$ since no prior states are known.}
    \label{fig:ar-denseEDpred}
\end{figure}

\noindent
The prediction of time series is a classical machine learning problem with many models specifically designed for such tasks most predominately seen in the field of neural language processing.
Among the most classical methods are standard recurrent neural networks which tend to be difficult to train~\cite{pascanu2013difficulty}, as well as long-short term memory (LSTM) architectures~\cite{hochreiter1997long}.
Recent advances of modeling time series include hierarchical networks~\cite{du2015hierarchical}, attention networks~\cite{kim2017structured} and transformer networks~\cite{vaswani2017attention}.
For modeling dynamical systems, we propose the following auto-regressive dense encoder-decoder model (AR-DenseED) illustrated in Fig.~\ref{fig:ar-denseED} which does not rely on any latent variable recurrent connections or specific gate design as seen in LSTMs~\cite{hochreiter1997long}.
The key philosophy of AR-DenseED is to efficiently model time integration by learning how to evolve the system forward in time given $k+1$ previous states.
In a true auto-regressive nature, the model predicts the dynamics of the system through sequential forward passes using the previous predictions as inputs as outlined in Algorithm~\ref{algo:testing}.
This is shown in Fig.~\ref{fig:ar-denseEDpred} where an AR-DenseED using three previous time-steps as inputs predicts five time-steps into the future.
\begin{remark}
    The number of previous time-steps used in the model's input, $\bm{\chi}^{n+1} \equiv \left\{\bm{u}^{n}, \bm{u}^{n-1}, \ldots, \bm{u}^{n-k} \right\}$, is a tunable hyper-parameter that can be adjusted depending on the system of interest.
    For all of the numerical examples tested, we found that including multiple time-steps in the input (i.e.  $k \ge 1$) was essential for improving training stability.
    However, using too many past time-steps in $\bm{\chi}$ slows training and does not yield considerable predictive improvements.
\end{remark}
\begin{algorithm}
    \caption{AR-DenseED Prediction}
    \label{algo:testing}
    \KwIn{Trained neural network model: $f\left(\cdot,\textbf{w}\right)$; Test initial state: $\bm{u}_{0}$; Max number of time-steps to predict: $T_{max}$}
    $\bm{\chi}^{1} \leftarrow \left\{\bm{u}_{0}, \bm{u}_{0}, \ldots, \bm{u}_{0} \right\}; \quad \bm{u}_{out}[0]=\bm{u}_{0}$ \;
    \For{$n = 1$ \KwTo $T_{max}$}{
        $\bm{u}^{n} \leftarrow f\left(\bm{\chi}^{n},\textbf{w}\right)$ \Comment*[r]{Forward pass of model}
        $\bm{u}_{out}[n] \leftarrow \bm{u}^{n}$\;
        % $\bm{\chi}^{n+1} \leftarrow \left\{\bm{u}^{n}, \bm{u}^{n-1}, \ldots, \bm{u}^{n-k}\right\}$\Comment*[r]{Update input}
        $\bm{\chi}^{n+1} \leftarrow \left\{\bm{u}^{n}, \bm{\chi}^{n}[0], \bm{\chi}^{n}[1], \ldots, \bm{\chi}^{n}[k-1]\right\}$ \Comment*[r]{Update input}
    }
    \KwOut{Predicted time series $\bm{u}_{out}=\left[\bm{u}_{0}, \bm{u}^{1}, \bm{u}^{2}, ..., \bm{u}^{T_{max}}\right]$;}
\end{algorithm}

AR-DenseED is built using successive layers of encoding/decoding convolutions and dense blocks originally proposed for solving elliptic systems in~\cite{zhu2018bayesian, zhu2019physics}.
The convolutional encoder-decoder can be parameterized as the function composition:
\begin{gather}
    f\left(\bm{\chi}^{n},\textbf{w}\right) = decoder \circ encoder(\bm{\chi}^{n}),\\
    \bm{u}^{n} = decoder\left(\bm{z}, \textbf{w}_{d}\right), \quad \bm{z} = encoder\left(\bm{\chi}^{n}, \textbf{w}_{e}\right),
\end{gather}
where $\left\{\textbf{w}_{e}, \textbf{w}_{d}\right\} = \textbf{w}$ are the encoder and decoder parameters, respectively.
$\bm{z}$ are latent variables that are of lower dimensionality than $\bm{u}^{n}$.
The input channel dimensionality of the convolutional model is the product of the number of state-variables $d_{0}$ and the number of previous time-steps used in $\bm{\chi}^{n}$.
The number of output channels is equivalent to the number of state-variables that are predicted at a given time-step.
While we use $f\left(\bm{\chi}^{n},\textbf{w}\right)$ to encapsulate this process, one can interpret this model as learning two processes: encoding data from previous time-steps to a latent variable $\bm{z}$ and the prediction of the next state as a function of $\bm{z}$.
We choose to use convolutional neural networks largely because they have been shown to yield faster convergence and better predictive capability for physics-constrained training compared to fully connected networks~\cite{zhu2019physics}.
Additionally, convolutional neural networks   require significantly less learnable weights than fully connected networks due to parameter sharing which allows for faster predictions~\cite{goodfellow2016deep}.
Computational efficiency during test time is imperative for surrogate modeling where prediction speed of the surrogate needs to outperform a numerical solver to justify its use.
Extensions of convolutional neural networks to unstructured and non-Euclidean domains can be achieved through geometric deep learning~\cite{bronstein2017geometric}, an emerging field that focuses on extending convolutional operators past structured data.
This will clearly require approximating the   spatial gradients of the PDE   on unstructured grids.

\subsection{Physics-Constrained Loss Function}
\label{sec:ar-denseed-loss}
\noindent
To train this model, we will be extending the previous work of Zhu~\etal~\cite{zhu2019physics} where governing PDEs are used to formulate a loss function.
The model is trained such that its predictions satisfy the governing equations of the system requiring only initial states, $\bm{u}_{0}$.
For clarity, we will refer to these initial states as \textit{training scenarios}.
Unlike works that have used fully connected networks for learning PDEs solutions in a similar manner (e.g. \cite{meade1994numerical, lagaris1998artificial, raissi2019physics, karumuri2019simulator}), our convolutional neural network surrogate requires the 
gradients in the governing PDEs to be numerically approximated.
In past works regarding the surrogate modeling of elliptic PDEs~\cite{zhu2019physics}, finite-difference based approximations were used to compute spatial gradients.
These approximations were found to be very effective and computationally efficient.
Thus a similar approach will be used for spatial derivatives in this work.

In dynamical systems this leaves one more critical component: the time-integrator or time-stepping method~\cite{ralston2001first}.
As previously outlined in~\Eqref{eq:general-loss}, we will pose the optimization of this model for a given time-step in terms of minimizing the difference between the model's predictions and a discrete numerical time-integrator $T_{\Delta t}$ of the governing PDE.
The standard $L_{2}$ loss for a series of $N$ time-steps and a mini-batch of $M$ training scenarios, $\mathcal{S}=\left\{\bm{u}_{0,i}\right\}_{i=1}^{M}$, can be posed as follows:
\begin{equation}
    \mathcal{L} = \frac{1}{M}\sum_{j=1}^{M}\sum_{i=1}^{N}\norm{\hat{\bm{u}}_{j}^{i} - \bm{u}_{j}^{i}}^{2}_{2}, \quad \hat{\bm{u}}_{j}^{i} = T_{\Delta t}\left(\bm{\mathcal{U}}_{j}^{i}, F_{\Delta x}\right),
    \label{eq:time-int-loss}
\end{equation}
where $\bm{u}^{i}$ is the prediction from the neural network $f\left(\bm{\chi}^{i}, \textbf{w}\right)$ making the loss implicitly dependent on all states within $\bm{\chi}^{i}$ and $\hat{\bm{u}}^{i}$ is the ``target'' calculated using the numerical time-integrator.
The $L_{2}$ norm corresponds to a finite integral over the entire domain $\Omega$.
It is  important to recognize that the discretization of both the time and spatial derivatives has introduced numerical truncation error into our loss function.
These errors in the deterministic case will ultimately be neglected, however we will expand on the idea of numerical error in the Bayesian model in Section~\ref{sec:bar-denseed}.

This formulation allows for  any time integration algorithm to be used, making it very versatile.
Thus one can select a numerical integrator that has the desired properties regarding stability, computational cost and accuracy.
In this work, we are interested in predicting large time-steps or when the model represents a time-step, $\Delta t$, resulting in a Courant-Friedrichs-Lewy (CFL) number greater than one.
In this regime, explicit methods are fundamentally unstable for hyperbolic PDEs thus one must resort to costly implicit methods which require expensive matrix inversions~\cite{ralston2001first, leveque2007finite}.
Rather than performing a matrix inversion as is the case for an implicit time integration scheme, the neural network's predictions for the next time-step are used when evaluating the spatial gradients in $T_{\Delta t}$.
This allows for an implicit like time integration without the need for matrix inversions during optimization.
Alternatively, $T_{\Delta t}$ could encapsulate multiple explicit calculations each at smaller time-steps. 
This strategy was not investigated in   this work but could allow for greater accuracy at the cost of additional computation.
\begin{remark}
    The use of discretization methods makes the model vulnerable to the same numerical issues that plague 
    each numerical approximation.
    Specifically with regards to the time-integrator, while an implicit scheme may be stable for very large time-steps this comes with the implications of reduced accuracy which should be considered.
    However, the parametrization of the solution as a neural network potentially relaxes the traditional numerical analysis constraints regarding stability and accuracy. 
    Thus this model could allow for large time-step predictions at a greater precision than vanilla numerical methods.
\end{remark}

A point that remains to be seen is if this optimization function will lead to the solution of the PDE. 
By substituting the time-integrator $T_{\Delta t}$ into the loss, one can arrive at the following proposition:
\begin{prop}
    The $L_2$ minimization between the model's prediction and a \textit{consistent} numerical time integration method is analogous to the $L_2$ minimization of the discretized PDE residual.
\end{prop}
\noindent
Intuitively, this is a logical statement since time integration methods are formulated on discretizing the temporal derivative.
We can easily show this with a simple example for a single time-step from $n\rightarrow n+1$. Consider the standard forward Euler time integration scheme:
\begin{align}
    \frac{\hat{\bm{u}}^{n+1}-\bm{u}^{n}}{\Delta t} &= - F_{\Delta x}\left(\bm{x}, \bm{u}^{n}\right)\\
    \hat{\bm{u}}^{n+1} = T_{\Delta t}\left(\bm{\mathcal{U}}^{n+1}, F_{\Delta x}\right)  &= \bm{u}^{n} - \Delta t\, F_{\Delta x}\left(\bm{x}, \bm{u}^{n}\right),
\end{align}
where $\bm{\mathcal{U}}^{n+1}=\left\{\bm{u}^{n}\right\}$.
Substituting this into the loss function in~\Eqref{eq:time-int-loss}:
\begin{align}
    \mathcal{L} &= \frac{1}{M}\sum_{i=1}^{M}\norm{\bm{u}_{i}^{n} - \Delta t\, F_{\Delta x}\left(\bm{x}_{i},\bm{u}^{n}_{i}\right) - \bm{u}_{i}^{n+1}}^{2}_{2}, \nonumber\\
    &= \frac{1}{M}\sum_{i=1}^{M}\norm{\bm{u}_{i}^{n+1} - \bm{u}_{i}^{n} + \Delta t\, F_{\Delta x}\left(\bm{x}_{i},\bm{u}^{n}_{i}\right)}^{2}_{2}, \nonumber\\
    &= \frac{\Delta t^{2}}{M}\sum_{i=1}^{M}\norm{\frac{\bm{u}_{i}^{n+1} - \bm{u}_{i}^{n}}{\Delta t} + F_{\Delta x}\left(\bm{x}_{i},\bm{u}^{n}_{i}\right)}^{2}_{2}, \nonumber\\
    &= \frac{\Delta t^{2}}{M}\sum_{i=1}^{M}\norm{\mathcal{R}\left(\bm{u}_{i}^{n+1}\right)}^{2}_{2} = \frac{\Delta t^{2}}{M}\sum_{i=1}^{M}\norm{\mathcal{R}\left(f\left(\bm{\chi}_{i}^{n+1}, \textbf{w}\right)\right)}^{2}_{2},\label{eq:resid-derivation}
\end{align}
we arrive at the minimization of the residual, $\mathcal{R}$, for the discretized PDE given the neural network's prediction.
The same result can be obtained for implicit time integration methods as well.
For example, this can easily be seen for the implicit backward Euler scheme by replacing $F_{\Delta x}\left(\bm{x}_{i},\bm{u}^{n}_{i}\right)$ in~\Eqref{eq:resid-derivation} with $F_{\Delta x}\left(\bm{x}_{i},\bm{u}^{n+1}_{i}\right)$ where $\bm{u}^{n+1}_{i}$ is the model's prediction.
Thus, if we use a time integration method that is consistent with the governing PDE, this optimization function minimizes the residual of the PDE across the entire domain leading to the solution of the discretized PDE.
After all, the minimization of the residual is the foundation for many iterative numerical methods (e.g. \cite{saad1986gmres, patankar1983calculation}) and has been empirically shown to be very effective for learning PDEs with deep learning models in past works~\cite{zhu2019physics, lagaris1998artificial, raissi2019physics}.

To enforce boundary conditions, one can introduce an additional loss term that enforces the desired response at the domain boundary.
This has been successfully implemented in past work for elliptic PDEs~\cite{zhu2019physics} and can easily be generalized to dynamical problems.
The systems in this work all have periodic boundaries which are enforced by using circular padding in PyTorch~\cite{paszke2017automatic} for all model convolutions as well as when evaluating the loss.
This is the equivalent to the use of ``ghost nodes'' in numerical simulations for enforcing periodicity.
Additional loss terms can be added to impose other prior physical constraints or conservation laws (\textit{e.g.} solenoidality or mass conservation).
Alternatively, one can modify the network architecture directly to impose constraints such as positivity and invariances~\cite{geneva2019quantifying, ling2016reynolds}.

\subsection{AR-DenseED Training}
\begin{figure}[H]
    \centering
    \includegraphics[width=0.9\textwidth]{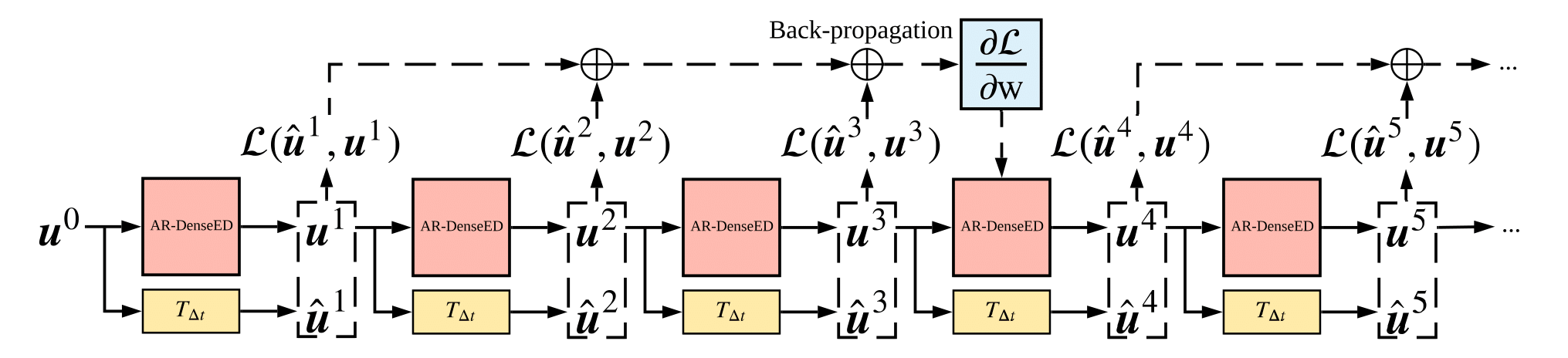}
    \caption{Multi-time-step back-propagation of the AR-DenseED where $\textbf{w}$ are the model's learnable weights, $\bm{u}^{n}$ is the model's prediction at time-step $n$, $\hat{\bm{u}}^{n}$ is the target value calculated using the numerical time-integrator $T_{\Delta t}$, and $\mathcal{L}$ is the physics-constrained loss at a single time-step.
    In this example, the computational graph evaluated during back-propagation spans all three predictions resulting in each contributing to the gradient descent update.}
    \label{fig:multi-backprop}
\end{figure}

\noindent
This model has several important advantages that we will leverage.
The first is that time is not an explicit input, thus this model has no fundamental limitation on its predictive range or initial conditions making it easier to train and better at extrapolation.
This is a key advantage of this model compared to the standard fully-connected networks that use time as a discrete input which fundamentally limits their predictive capabilities~\cite{lagaris1998artificial,raissi2019physics, raissi2018deep}.
During training, the model is allowed to progressively explore the system and learn the dynamics by slowly ``unrolling'' the number of time-steps it predicts as training progresses.
For example, at the beginning of training, the model only predicts a few time-steps from its initial state, however this may increase to hundreds of time-steps as training continues.
This allows the model to discover and learn dynamics that may be absent from the provided initial states.
Additionally, since the model's output prediction is then the input for the next time-step, we can back-propagate through multiple time-steps as illustrated in Fig.~\ref{fig:multi-backprop} without latent variable recurrent connections.
Allowing the model to auto-regress itself forward in time and compute back-propagation through multiple time-steps promotes the learning of continuous time series.
In practice, we only back-propagate through a small number of time-steps to avoid vanishing gradient issues.
The training process is outlined in Algorithm~\ref{algo:training}.
\begin{algorithm}
    \caption{Training AR-DenseED}
    \label{algo:training}
    \KwIn{Neural network model: $f\left(\cdot,\textbf{w}\right)$; Back-prop interval: $p$; Max number to time-steps to unroll: $T_{max}$; Number of epochs: $N$; Learning rate: $\eta$}
    $\textrm{tsteps} = \textrm{linspace}\left(p, T_{max}, N\right)$ \;
    \For{$\textrm{epoch} = 1$ \KwTo $N$}{
        $\bm{\chi}^{1} \leftarrow \left\{\bm{u}_{0}, \bm{u}_{0}, \ldots, \bm{u}_{0} \right\}$ \;
        $T \leftarrow \textrm{tsteps}[\textrm{epoch}]$ \Comment*[r]{Time-steps to unroll}
        \For{$i = 1$ \KwTo $T$}{
            $\bm{u}^{i} \leftarrow f\left(\bm{\chi}^{i},\textbf{w}\right)$ \Comment*[r]{Forward pass of the model}
            $\hat{\bm{u}}^{i} = T_{\Delta t}\left(\bm{\mathcal{U}}^{i}, F_{\Delta x}\right)$ \Comment*[r]{Time integration}
            $\mathcal{L}^{i}$ = $\mathcal{L}^{i-1}$ + $\textrm{MSE}\left(\hat{\bm{u}}^{i}, \bm{u}^{i}\right)$ \Comment*[r]{Calculate Loss}
            \If{Mod(n,p)=0}{ 
                $\nabla \textbf{w} \leftarrow  \textrm{Backprop}(\mathcal{L}^{i})$ \Comment*[r]{Multi-step back-prop.}
                $\textbf{w} \leftarrow \textbf{w} - \eta \nabla \textbf{w}$ \Comment*[r]{Gradient Descent}
                 $\mathcal{L}^{i} = 0$ \Comment*[r]{Zero loss}
            }
            $\bm{\chi}^{i+1} \leftarrow \left\{\bm{u}^{i}, \bm{\chi}^{i}[0], \bm{\chi}^{i}[1], \ldots, \bm{\chi}^{i}[k-1]\right\}$ \Comment*[r]{Update input}
        }
    }
    \KwOut{Trained auto-regressive model $f\left(\cdot,\textbf{w}\right)$;}
\end{algorithm}

\section{Bayesian AR-DenseED}
\label{sec:bar-denseed}
\noindent
A challenge of physics-constrained learning with no output training data is developing a meaningful probabilistic framework.
In past works, a probabilistic surrogate was proposed using the Boltzmann distribution as a reference density and minimizing the Kullback-Leibler divergence for a generative model~\cite{zhu2019physics,yang2018adversarial}.
While such methodologies represent some built in uncertainty in the model and can yield reasonable error bars, the true interpretation of the resulting uncertainty is much less concrete.
Thus in this work, we propose a novel Bayesian framework for physics-constrained models that allow for interpretable uncertainty measures to be produced in the absence of training data.

To formulate the Bayesian AR-DenseED (BAR-DenseED) model, we wish to account for the two major uncertainty components: \textit{aleatoric} uncertainty which quantifies noise in the observations and 
\textit{epistemic} uncertainty which captures inherit uncertainty in the model~\cite{kendall2017uncertainties}.
Epistemic uncertainty is associated with the confidence of the model's predictions which is influenced by factors such as limited training scenarios, limited expressibility of the model, etc.
For DNNs, epistemic uncertainty is most commonly captured by placing priors on the parameters of the model often being formulated as a Bayesian neural network~\cite{neal2012bayesian}.
Aleatoric uncertainty involves the noise that potentially exists in the data on which the model is trained on, and is often captured by placing a distribution over the model's outputs~\cite{gal2016uncertainty}.
In a data-driven sense, aleatoric uncertainty arises from the simulators or sensors used to collect the training data.
In the physics-constrained learning paradigm, we will interpret aleatoric uncertainty as the quantification of error associated with the truncation error introduced when formulating the physics-constrained loss function.
As discussed in Section~\ref{sec:ar-denseed-loss}, the physics-constrained optimization of the auto-regressive model is posed as the minimization of the error between the model's predictions and a numerical time-integrator of the same time-step size.
However, this numerical time-integrator introduces truncation error:
\begin{equation}
    \bm{u}_{i}^{n+1} = T_{\Delta t}\left(\bm{\mathcal{U}}^{n+1}, F_{\Delta x}\right) + E_{\Delta t} + E_{\Delta x},
\end{equation}
where $E_{\Delta t}$ and $E_{\Delta x}$ denote the error associated with the discretization of the temporal and spatial derivatives, respectively.
In the deterministic case, such errors are neglected, however, depending on the resolution of the spatial discretization or the time-step size these errors can impact a numerical solver's accuracy.
For most numerical solvers using explicit time integration schemes, the bulk of this error arises from the discretization of the spatial derivatives.
Alternatively, when predicting large time-steps with implicit methods the discrete time integration can become the primary source of error due to numerical diffusion~\cite{iserles2009first}.

\subsection{Posterior Formulation}
\noindent
With an idea of the sources of uncertainty we wish to account for, let us start with defining a posterior over the model parameters.
In a data-driven model, the   likelihood captures the probability of the observations for a given model.
For physics-constrained learning where no observations are available, we take as  ``target data''  the prediction of the numerical time-integrator $\hat{\bm{u}}$.
Similar to data-driven probabilistic models~\cite{zhu2018bayesian,geneva2019quantifying}, we account for the potential discretization error that may arise from $T_{\Delta t}$ through additive output-wise noise, $\epsilon$, for a single arbitrary time-step:
\begin{equation}
    \hat{\bm{u}}^{i} = f\left(\bm{\chi}^{i},\textbf{w}\right) + \epsilon, \quad p\left(\epsilon\right)= \mathcal{N}\left(\epsilon\,|0,\beta^{-1}\bm{I}_{d}\right),
\end{equation}
where for mathematical convenience, we represent the discretized state variables $\bm{u}$ and $\hat{\bm{u}}$ as vectors in $\mathbb{R}^{d}$.
$\bm{I}_{d}$ denotes the identity matrix in $\mathbb{R}^{d\times d}$.
The additive noise is taken as Gaussian with a \textit{learnable} precision $\beta$.
Thus the likelihood for a single time-step, $i$, becomes the following:
\begin{align}
   p\left(\hat{\bm{u}}^{i}|\bm{\chi}^{i},\textbf{w}, \beta\right)&=\mathcal{N}\left(\hat{\bm{u}}^{i}|f\left(\bm{\chi}^{i},\textbf{w}\right), \beta^{-1}\bm{I}_{d}\right), \nonumber\\
   &=\mathcal{N}\left(T_{\Delta t}\left(\bm{\mathcal{U}}^{i}, F_{\Delta x}\right)|f\left(\bm{\chi}^{i},\textbf{w}\right), \beta^{-1}\bm{I}_{d}\right).
\end{align}
Both the inputs to the model $\bm{\chi}^{i}$  and time-integrator $\bm{\mathcal{U}}^{i}$ are found from the deterministic evolution of the model until time-step $i$.
Under the Markov assumption, we can formulate the likelihood of an entire time-sequence as the product of individual steps:
\begin{align}
    p\left(\hat{\bm{u}}^{N}, \ldots, \hat{\bm{u}}^{1}|\bm{u}_{0},\textbf{w}, \beta\right) &= \prod_{i=1}^{N}p\left(\hat{\bm{u}}^{i}| \bm{\chi}^{i},\textbf{w}, \beta\right), \nonumber \\
    &= \prod_{i=1}^{N}\mathcal{N}\left(T_{\Delta t}\left(\bm{\mathcal{U}}^{i}, F_{\Delta x}\right)| f\left(\bm{\chi}^{i},\textbf{w}\right), \beta^{-1}\bm{I}_{d}\right). \label{eq:likelihood}
\end{align}
Note that the number of past time-steps contained in the model's input, $\bm{\chi}^{i}$, is analogous to the order of the Markov chain.
In this likelihood, the numerical time-integrator, $T_{\Delta t}$, can be interpreted as calculating the target $\hat{\bm{u}}^{i}$ on-the-fly. 
Thus the evaluation of this likelihood function is in fact still data-less.
\begin{remark}
    To find the maximum likelihood estimate (MLE), minimization of the negative log likelihood is often taken as the optimization objective~\cite{bishop2006machine}.
    In this likelihood formulation, when minimizing the negative log likelihood in~\Eqref{eq:likelihood}, one recovers the standard $L_{2}$ loss as previously shown in~\Eqref{eq:time-int-loss} for the deterministic model.
    Thus the MLE is equivalent to the minimization of the strong residual of the discretized PDE for both implicit and explicit time integration schemes indicating the appropriateness of the 
    selected likelihood   for our physics-constrained model.
\end{remark}
A gamma prior is assigned to the noise precision $\beta$:
\begin{equation}
    p(\beta)=\Gamma\left(\beta|a_{1},b_{1}\right),
\end{equation}
where $a_{1}$ and $b_{1}$ are the shape and rate parameters, respectively.
The hyper-parameters of the $\beta$ prior are set based on the \textit{a priori} estimate of the magnitude of the discretization error of $T_{\Delta t}$ and  $F_{\Delta x}$.
Since we intend to build large time-step surrogate models with $CFL > 1$, we will assume that the majority error arises from the time integration method.
However, the following procedure can easily be extended to account for spatial discretization error as well.

Given an arbitrary system, we can express the temporal truncation error in the following formula which is standard in Richardson extrapolation~\cite{richardson1927viii}:
 \begin{align}
    \hat{\bm{u}}^{i} &= T_{\Delta t}\left(\bm{\mathcal{U}}^{i}, F_{\Delta x}\right) + c_{0}\left(\Delta t\right)^{k_{0}} + c_{1}\left(\Delta t\right)^{k_{1}} + c_{2}\left(\Delta t\right)^{k_{2}} + \dots \nonumber \\
     &= T_{\Delta t}\left(\bm{\mathcal{U}}^{i}, F_{\Delta x}\right) + c_{0}\left(\Delta t\right)^{k_{0}} + \mathcal{O}\left(\left(\Delta t\right)^{k_{1}}\right),\label{eq:richard0}
\end{align}
where $c_{i}$ are unknown constants, and $k_{i}$ are constants denoting the ``order'' of the error term such that $\left(\Delta t\right)^{k_{i}} > \left(\Delta t\right)^{k_{i+1}}$.
Under the assumption that higher-order terms are negligible, we take the expected value of our prior to be $\mathbb{E}\left(\beta^{-1}\right)=c_{0}\left(\Delta t\right)^{k_{0}}$. 
Additionally this prior is given a large variance to reduce its strength.
The parameters $c_{0}$ and $k_{0}$ can be estimated based on the magnitude of the state variables and the order of accuracy of the temporal discretization, respectively~\cite{oliver2014estimating}.
In this work, $c_{0}$ is set to be approximately $20\%$ the maximum value of the quantity of interest and $k_{0}$ is set to three which over estimates the accuracy of the second-order accurate time integration method used.
This encourages the model to be more accurate at the start of training rather than attributing initial prediction discrepancies to noise.

As is standard for Bayesian neural networks, the network's $K$ learnable parameters, $\textbf{w}$, are treated as random variables.
Due to the large number of weights in our model, we propose a fully factorizable zero mean Gaussian with a Gamma-distributed precision scalar $\alpha$:
\begin{equation}
    p\left(\mathbf{w}|\alpha\right)=\mathcal{N}\left(\mathbf{w}|0, \alpha^{-1}\bm{I}_{K}\right), \quad p\left(\alpha\right) = Gamma\left(\alpha|a_{0},b_{0}\right),
\end{equation}
where the rate parameter $a_{0}$ and the shape parameter $b_{0}$ are $0.5$ and $10$,  respectively.
This results in a prior with a Student's $\mathcal{T}$ density centered at zero that has a wider support region than a standard Gaussian.
In our past works~\cite{zhu2018bayesian, geneva2019quantifying}, a narrow Student's $\mathcal{T}$-distribution was used to more strongly promote sparsity~\cite{tipping2001sparse}, however it was found that a narrow prior would damage the 
predictive capability of BAR-DenseED.
Thus the sparsity requirement is relaxed while still regulating the magnitude of the model's weights.
We note that when one uses an optimizer with momentum and weight decay, such as ADAM~\cite{kingma2014adam}, an implicit prior on the weights is enforced which is largely ambiguous~\cite{loshchilov2018decoupled}.
Since we will ultimately approximate the posterior, this prior does not need to be accounted for in the formulation of the joint posterior used for optimization.
As a result for a batch of $M$ i.i.d. training scenarios, $\mathcal{S}=\left\{\bm{u}_{0,i}\right\}_{i=1}^{M}$, the posterior of the network is as follows:
\begin{align}
    p\left(\textbf{w},\beta|\mathcal{S}\right) & \sim \prod^{M}_{j=1} p\left(\hat{\bm{u}}_{j}^{N}, \ldots, \hat{\bm{u}}_{j}^{1}|\bm{u}_{0,j},\textbf{w},\beta\right)p\left(\textbf{w}\right)p\left(\beta\right),  \nonumber \\ 
    & \sim \prod^{M}_{j=1}\prod_{i=1}^{N}\left[p\left(\hat{\bm{u}}_{j}^{i}|\bm{\chi}_{j}^{i},\textbf{w},\beta\right)\right]p\left(\textbf{w}|\alpha\right)p\left(\alpha|a_{0},b_{0}\right)p\left(\beta|a_{1},b_{1}\right),  \nonumber\\
    & \begin{multlined}\sim \prod^{M}_{j=1}\prod_{i=1}^{N} \left[\mathcal{N}\left(T_{\Delta t}\left(\bm{\mathcal{U}}_{j}^{i}, F_{\Delta x}\right)| f\left(\bm{\chi}_{j}^{i},\textbf{w}\right), \beta^{-1}\bm{I}_{d}\right)\right] \mathcal{N}\left(\mathbf{w}|0, \alpha^{-1}\bm{I}_{K}\right)\\
    \qquad \Gamma\left(\alpha| a_{0}, b_{0}\right) \Gamma\left(\beta|a_{1},b_{1}\right).
    \end{multlined}
    \label{eq:posterior-form}
\end{align}
\begin{remark}
    The task of computing the maximum \textit{a posteriori} probability (MAP) estimate  is closely related to   maximizing the likelihood with the addition of 
    appropriate weight regularization that arises from the use of priors on the model's parameters~\cite{bishop2006machine}. Thus, the MAP estimation 
    minimizes a regularized form of the previously considered  $L_{2}$ deterministic loss function that was defined based on the discretized PDE residual.
\end{remark}

\subsection{Posterior Approximation}
\noindent
The Bayesian paradigm seeks to represent model uncertainty by marginalizing out the model's parameters which results in the predictive distribution.
This marginalization is often not analytically tractable and is usually approximated with Monte Carlo sampling of  the posterior.
In earlier works,   Bayesian DNNs focused on the use of Monte Carlo or ensemble based methodologies~\cite{richard1991neural, barber1998ensemble, neal2012bayesian}.
However, with the number of parameters in such models growing exponentially larger over recent years, such traditional methods are  computationally intractable.
As a result,  many  recent   Bayesian deep learning frameworks focus on variational methods that fit a proposal distribution over the true posterior of the model's parameters.
Variational ideas have led to multiple   developments including: Bayes by back-prop~\cite{blundell2015weight}, Bayesian dropout approximation~\cite{gal2016dropout} and Stein variational gradient descent~\cite{liu2016stein}.

For sampling the posterior of BAR-DenseED, we will use a recently proposed Stochastic Weight Averaging Gaussian (SWAG)~\cite{maddox2019simple}.
SWAG is an approximate Bayesian method that builds upon the Stochastic Weight Averaging (SWA), an optimization methodology where running averages of model parameters are kept during the stochastic gradient descent (SGD) procedure~\cite{ruppert1988efficient, polyak1992acceleration, izmailov2018averaging}.
SWAG approximates the posterior in two phases:
\begin{enumerate}
    \item The model of interest is first trained using traditional machine learning methods to minimize the negative log of the posterior defined in~\Eqref{eq:posterior-form} (equivalent to solving for the MAP estimate).
    Specifically in this work, we optimize the model using the ADAM~\cite{kingma2014adam}, an extension of SGD, with exponential learning rate decay.
    \item Once the model has been trained, SGD is ran again at a constant learning rate.
    During this process, samples of the model's parameters are collected.
    The core idea is to use SGD to explore the local support region of the MAP estimate.
    These SGD iterations can provide useful information about the form of the posterior which can then be used to approximate the  posterior density function for full Bayesian inference.
\end{enumerate}
In SWAG, the posterior over the BAR-DenseED parameters is approximated as a Gaussian distribution with $S$ samples of the model's parameters:
\begin{equation}
    p(\bm{\theta}|\mathcal{S}) \sim \mathcal{N}\left(\bm{\theta}_{SWA}, \bm{\Sigma}_{SWA}\right), \quad \bm{\theta} \equiv \left\{\textbf{w}, ln(\beta)\right\},
\end{equation}
which is standard when using the Laplace approximation.
We note that the noise precision, $\beta$, has a log-normal posterior approximation to ensure that it is positive.
The mean and the covariance are approximated using the model parameters proposed by SGD:
\begin{gather}
    \bm{\theta}_{SWA}=\frac{1}{S}\sum_{i=1}^{S}\bm{\theta}_{i}, \quad \bm{\Sigma}_{SWA}=\frac{1}{2}\left(\bm{\Sigma}_{Diag}+\bm{\Sigma}_{lr}\right),\\
    \bar{\bm{\theta}}^{2}=\frac{1}{S}\sum_{i=1}^{S}\bm{\theta}^{2}_{i}, \quad \bm{\Sigma}_{Diag} = \textrm{Diag}\left(\bar{\bm{\theta}^{2}}-\bm{\theta}_{SWA}^{2}\right),\\
    \bm{\Sigma}_{lr}=\frac{1}{K-1}\bm{D}\bm{D}^{T}, \quad \bm{D}_{i} = \left(\bm{\theta}_{i}-\bm{\theta}_{SWA}\right),
\end{gather}
where $\bm{\theta}_{i}$ are the model parameters at epoch $i$, and $\bm{D} \in \mathbb{R}^{K\times H}$ is a deviation matrix consisting of the $H$ most recent parameter samples forming a low-rank approximation.
$\bm{D}_{i} \in \mathbb{R}^{K}$ is a column of this deviation matrix.
This results in a simple sampling method outlined in Algorithm~\ref{algo:swag} approximating the posterior of the model parameters for a time series prediction.
While many other Bayesian approaches exist, we selected to use SWAG for three main reasons: its simplicity in both formulation and implementation, its non-invasive nature to the learning of the neural network and its low computational overhead compared to other Bayesian approaches.
These factors are important for the learning of time series problems since learning generally becomes significantly more expensive and difficult.
\begin{algorithm}
    \caption{Approximating the BAR-DenseED Posterior with SWAG~\cite{maddox2019simple}}
    \label{algo:swag}
    \KwIn{Pre-trained model parameters optimized for MAP: $\bm{\theta}_{0}$; Number epochs to run: $N$; Time series training length: $T$; Sample frequency: $p$; Size of low-rank approximation matrix: $H$; Learning rate: $\eta_{swag}$; Negative log posterior: $\mathcal{L}_{p}$}
    $\bar{\bm{\theta}}=\bm{\theta}_{0}, \quad \bar{\bm{\theta}^{2}}=\bm{\theta}_{0}^{2}$\;
    $n=1$ \Comment*[r]{Number of sampled models}
    \For{$i = 1$ \KwTo $N$}{
        \For{$j = 1$ \KwTo $T$}{
        $\bm{u}^{j} \leftarrow f\left(\bm{\chi}^{j},\textbf{w}\right)$ \Comment*[r]{Forward pass of the model}
        $\hat{\bm{u}}^{j} = T_{\Delta t}\left(\bm{\mathcal{U}}^{j}, F_{\Delta x}\right)$ \Comment*[r]{Time integration}
        $\bm{\theta}_{j} = \bm{\theta}_{j-1} - \eta_{swag}\nabla_{\bm{\theta}}\mathcal{L}_{p}\left(\hat{\bm{u}}^{j}, \bm{u}^{j}\right)$ \Comment*[r]{Multi-step back-prop.}
        $\bm{\chi}^{j+1} \leftarrow \left\{\bm{u}^{j}, \bm{\chi}^{j}[0], \bm{\chi}^{j}[1], \ldots, \bm{\chi}^{j}[k-1]\right\}$ \Comment*[r]{Update input}
        }
        \If{Mod(i,p)=0}{ 
            $\bar{\bm{\theta}} = \frac{n\bar{\bm{\theta}} + \bm{\theta}_{i}}{n+1}$ \Comment*[r]{First moment update}
             $\bar{\bm{\theta}^{2}} = \frac{n\bar{\bm{\theta}^{2}} + \bm{\theta}^{2}_{i}}{n+1}$\Comment*[r]{Second moment update}
             $\hat{\bm{D}}=\textrm{concat}\left(\hat{\bm{D}}[:,\,-(H-1):],\bm{\theta}_{i}, \: \textrm{dim}=1\right)$\;
             $n = n + 1$\;
        }
    }
    $\bm{D} = \hat{\bm{D}} - \bar{\bm{\theta}}$\Comment*[r]{Low-rank deviation matrix}
    \KwOut{$\bm{\theta}_{SWA}=\bar{\bm{\theta}}; \: \bm{\Sigma}_{Diag} = \bar{\bm{\theta}^{2}} - \bm{\theta}_{SWA}^{2}; \: \bm{D};$}
\end{algorithm}

Of significant importance is the sampling learning rate, which is directly related to both the shape as well as the convergence rate of the posterior.
As discussed in~\cite{maddox2019simple}, this learning rate should be large enough to sufficiently explore the support region of the minima the model has converged to.
However, $\eta_{swag}$ should not be too large such that the model potentially jumps to other local minima during the collection of samples to approximate the posterior.
This is due to the use of a single mode Gaussian as the approximate density for which multimodal data cannot be handled robustly.
\begin{remark}
    While vanilla stochastic gradient descent is shown here for  simplicity of illustrating the SWAG algorithm, fundamentally, one can use other optimization methods as well to collect SWAG samples such as gradient descent with momentum~\cite{maddox2019simple}.
    Similar to the use of various Markov chain Monte Carlo methods for approximating density functions, different stochastic optimization methods can be used to approximate the density of the posterior of the neural network.
\end{remark}

\subsection{Predictive Statistics}
\noindent
As previously discussed, predictive statistics in a Bayesian framework are obtained by marginalizing or integrating out the parameters of the model.
Given the previous Markov assumption, the predictive distribution can be posed in terms of a single arbitrary time-step.
We will approximate the marginalization of the model parameters, often referred to as Bayesian model averaging, using Monte Carlo with $P$ parameter samples:
\begin{align}
    p\left(\hat{\bm{u}}^{*}|\bm{\chi}^{*},\mathcal{S}\right)&=\int p\left(\hat{\bm{u}}^{*}|f\left(\bm{\chi}^{*},\textbf{w}\right) ,\beta^{-1}\bm{I}\right)p\left(\textbf{w},\beta|\mathcal{S}\right)d\textbf{w}d\beta,  \nonumber \\
    &\approx \frac{1}{P}\sum_{i=1}^{P}p\left(\hat{\bm{u}}^{*}|f\left(\bm{\chi}^{*},\textbf{w}_{i}\right) ,\beta_{i}^{-1}\bm{I}\right), \quad \left\{\textbf{w}_{i},\beta_{i}\right\} \sim p\left(\bm{\theta}|\mathcal{S}\right),
\end{align}
where $\bm{\chi}^{*}$ and $\hat{\bm{u}}^{*}$ are the predictive inputs and outputs, respectively.
The posterior, $p(\bm{\theta}|\mathcal{S}) \equiv p(\textbf{w}, ln(\beta)|\mathcal{S})$, has been approximate by SWAG and is easily sampled from.
The predictive expectation can be obtained as follows:
\begin{align}
    \mathbb{E}\left[\hat{\bm{u}}^{*}|\bm{\chi}^{*},\mathcal{S}\right]&=\mathbb{E}_{p\left(\bm{\theta}|\mathcal{S}\right)}\left[\mathbb{E}\left(\hat{\bm{u}}^{*}|\bm{\chi}^{*},\textbf{w},\beta\right)\right],  \nonumber  \\
    &=\mathbb{E}_{p\left(\textbf{w}|\mathcal{S}\right)}\left[f\left(\bm{\chi}^{*},\textbf{w}\right)\right] \approx \frac{1}{P}\sum_{i=1}^{P}f\left(\bm{\chi}^{*},\textbf{w}_{i}\right),
    \label{eq:predictive-mean}
\end{align}
where the additive output noise is not present due to its zero-mean Gaussian density.
The predictive conditional covariance can also be obtained in a similar fashion:
\begin{align}
    \textrm{Cov}\left[\hat{\bm{u}}^{*}|\bm{\chi}^{*},\mathcal{S}\right] &= \mathbb{E}_{p\left(\bm{\theta}|\mathcal{S}\right)}
    \left[\textrm{Cov}\left(\hat{\bm{u}}^{*}|\bm{\chi}^{*},\textbf{w},\beta\right)\right] + \textrm{Cov}_{p(\bm{\theta}|\mathcal{S})}\left(\mathbb{E}\left[\hat{\bm{u}}^{*}|\bm{\chi}^{*},\textbf{w},\beta\right]\right),  \nonumber \\
    &= \mathbb{E}_{p(ln(\beta)|\mathcal{S})}\left[\beta^{-1}\bm{I}\right] + \textrm{Cov}_{p(\textbf{w}|\mathcal{S})}\left(f\left(\bm{\chi}^{*},\textbf{w}\right)\right),  \nonumber \\
    &\approx \frac{1}{P}\sum_{i=1}^{P}\left[\beta^{-1}_{i}\bm{I}+f\left(\bm{\chi}^{*},\textbf{w}_{i}\right)f\left(\bm{\chi}^{*},\textbf{w}_{i}\right)^{T}\right]-\mathbb{E}\left[\hat{\bm{u}}^{*}|\bm{\chi}^{*},\mathcal{S}\right]\mathbb{E}\left[\hat{\bm{u}}^{*}|\bm{\chi}^{*},\mathcal{S}\right]^{T},
\end{align}
where $\mathbb{E}\left[\bm{u}^{*}|\bm{\chi}^{*},\mathcal{S}\right]$ has been defined in~\Eqref{eq:predictive-mean}.
To predict an entire time series, each model is sampled at the first time-step and  auto-regressed forward in time independently.
Each set of parameters sampled from the posterior can be interpreted as an individual particle that is propagated forward in time.
Thus as we predict further in time, we should expect the predictive variance of the model to gradually increase.

% ==== Kuramoto-Sivashinsky Equation ====
\section{Kuramoto-Sivashinsky Equation}
\label{sec:ks}
\begin{figure}[H]
    \centering
    \includegraphics[width=0.8\textwidth]{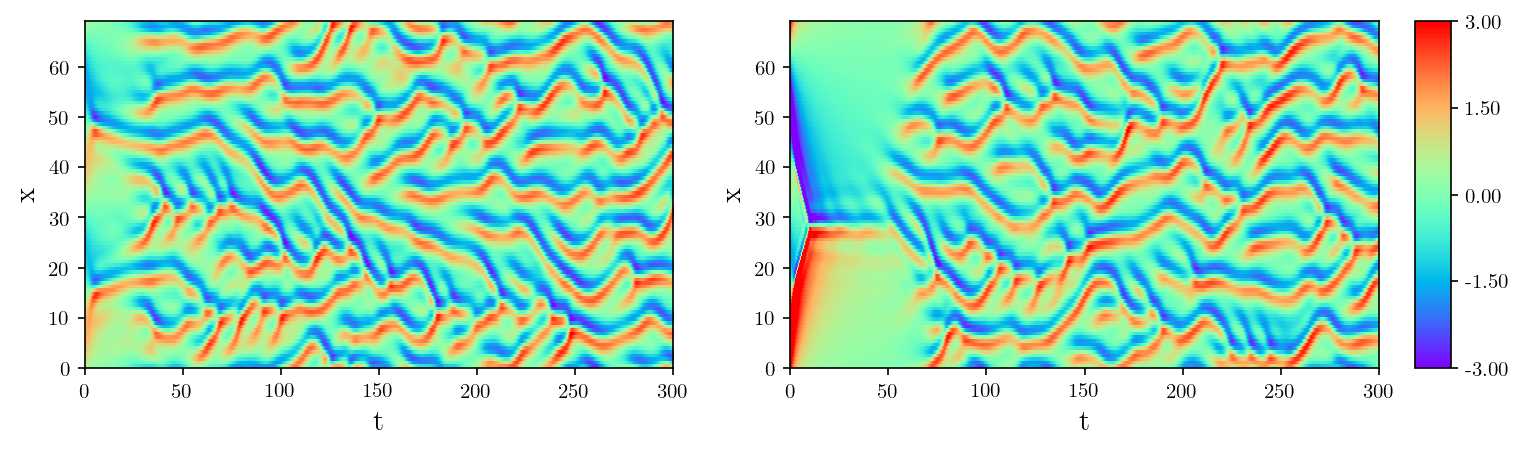}
    \caption{The Kuramoto-Sivashinsky equation for two different initial states solved using the spectral ETDRK4 scheme~\cite{cox2002exponential}.}
    \label{fig:ks-simulation}
\end{figure}

\noindent
The first physical system that we are interested in is the 1D Kuramoto-Sivashinsky (K-S) equation which is a fourth-order, nonlinear partial differential equation:
\begin{gather}
    \frac{\partial u}{\partial t} + u\frac{\partial u}{\partial x} + \frac{\partial^{2} u}{\partial x^{2}} + \viscosity\frac{\partial^{4} u}{\partial x^{4}}= 0,\\
    u(0,t) = u(L,t), \quad x\in[0,L], \quad t\in[0,T],
\end{gather}
where $\viscosity$ is referred to as the ``hyper-viscosity'' and is set to $\viscosity=1$ for the remainder of this section.
The K-S equation is widely known for its chaotic behavior when the size of the periodic domain is sufficiently large (generally $L\ge50$) in which the system becomes a spatio-temporally chaotic attractor~\cite{hyman1986kuramoto}.
The K-S PDE has attracted great interest as it serves as a prototypical problem for studying complex dynamics with its chaotic regime being weakly turbulent  (as opposed to strong turbulence seen in the Navier-Stokes equations)~\cite{hyman1986order, wittenberg1999scale}.
Several physical systems such as chemical phase turbulence, plasma ion instabilities and flame front instabilities have all seen the K-S equation arise within them~\cite{laquey1975nonlinear, kuramoto1976persistent, michelson1977nonlinear}.
For our problem of interest, we take the domain $L$ to be $[0,22\pi]$ putting the system well within its chaotic regime.
The domain is discretized by $96$ uniform cells and time-step is $\Delta t = 0.1$.
Two sample responses of the K-S equation for two different initial conditions are illustrated in Fig.~\ref{fig:ks-simulation}.
During training and testing, we will ignore the initial transient state thus our initial conditions will be already fully developed ``turbulence'' ($T\ge100$).

Our goal is for AR-DenseED to predict the chaotic response of the system accurately thus illustrating the potential of this model to predict physical dynamics.
In the past, others have attempted to model this system by machine learning methods.
Recently, in Pathak~\etal~\cite{pathak2017using}  reservoir computing was used to predict the K-S system, however the model is trained on the past history of a specific state.
Thus the model learns only for a specific initial condition.
The recent formulations of physics-informed neural networks in Raissi~\etal~\cite{raissi2019physics, raissi2018deep} have been able to work 
for learning a specific initial condition without training data, however these models have yet to be shown effective as a predictive surrogate.

The AR-DenseED used for the K-S equation consists of a single convolutional block, followed by a dense block, followed by a deconvolutional block resulting in a model with just over $4800$ learnable weights.
The two previous states of the system are used as inputs, $\bm{\chi}^{n+1}=\left\{\bm{u}^{n},\bm{u}^{n-1}\right\}$.
Similar to the numerical solver, the time-step size of the model is set to $\Delta t = 0.1$ with a spatial discretization of $96$ points.
As previously discussed, the negative log of the joint posterior in~\Eqref{eq:posterior-form} is the loss function.
For the physics-constrained loss function, the implicit Crank-Nicolson time integration is used and the remaining spatial gradients are discretized as follows:
\begin{gather}
    \begin{gathered}
    T_{\Delta t}(\bm{\mathcal{U}}^{n}, F_{\Delta x}) = \bm{u}^{n} + \Delta t \left[-0.5\left(F_{\Delta x}(\bm{x}, \bm{u}^{n+1}) + F_{\Delta x}(\bm{x}, \bm{u}^{n})\right)\right],\\ F_{\Delta x}(\bm{x}, \bm{u}^{n}) = u^{n}u^{n}_{x} + u^{n}_{xx} + u^{n}_{xxxx},\end{gathered}\\
    uu_{x} = \frac{-u^{2}_{i+2}+8u^{2}_{i+1}-8u^{2}_{i-1}+u^{2}_{i-2}}{24 \Delta x}, \\
    u_{xx} = \frac{-u_{i+2}+16u_{i+1}-30u_{i}+16u_{i-1}-u_{i-2}}{12\Delta x^{2}},\\
    u_{xxxx} = \frac{-u_{i+3}+12u_{i+2}-39u_{i+1}+56u_{i}-39u_{i-1}+12u_{i-2}-u_{i-3}}{6\Delta x^{4}},
\end{gather}
where the spatial gradients are approximated using fourth-order accurate finite difference discretizations that are implemented efficiently using convolutional operators.
The model was trained for $100$ epochs using $2560$ training scenarios that were generated using a truncated Fourier series with random coefficients discussed in~\ref{app:ks-initial}.
This Fourier series serves to approximate the physical turbulence of the system, and thus estimating the true distribution of the possible initial states $p(\bm{u}_{0})$.
During test time, we use $200$ test cases of true turbulent initial states of the system obtained from a numerical simulator.
This will demonstrate how one can use approximated training scenarios and physics-constrained learning to train a model that can be used on a true realization of the system.
The training scenarios were mini-batched with a batch size of $256$.
During training the model was allowed to unroll itself in time up to $1000$ time-steps to allow AR-DenseED to thoroughly explore the turbulent dynamics.
Training on a single 1080Ti GPU took approximately $1.5$ wall-clock hours.
Additional details on the model and training parameters are discussed in~\ref{app:ks}.

\subsection{AR-DenseED Deterministic Predictions}
\begin{figure}[H]
    \centering
    \includegraphics[width=0.9\textwidth]{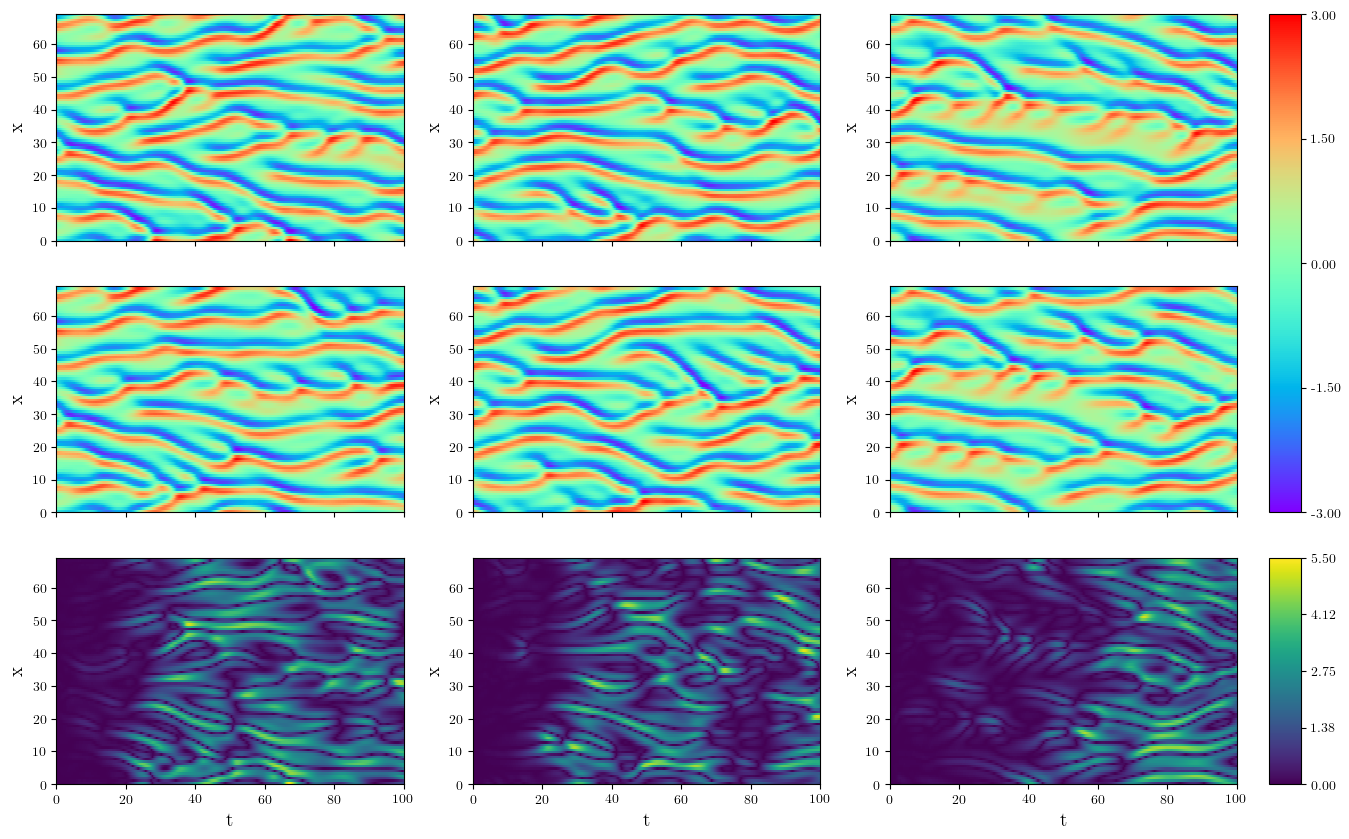}
    \caption{Three test predictions of the Kuramoto-Sivashinsky equation using AR-DenseED. (Top to bottom) Target field solved system using the spectral ETDRK4 scheme, AR-DenseED prediction and finally the $L_1$ error.}
    \label{fig:ks-test-prediction}
\end{figure}

\noindent
We start with the prediction of the deterministic AR-DenseED model.
Predictive results are shown  in Fig.~\ref{fig:ks-test-prediction} for three test initial conditions compared against a numerical solver.
All predictions are obtained by only providing the initial state and evolving the system with $1000$ consecutive iterations of the neural network.
Overall the results are very impressive and are significant improvements compared to past literature despite only using a single initial state to predict.
The model is able to maintain consistency with the numerical solver for between $t=[0,30]$, but then diverges due to small prediction error that causes the model to shift its response as a result of the systems' chaotic nature.
However, the predicted system remains qualitatively reasonable and stable for even extended times which is a significant advantage of our auto-regressive formulation.

\begin{figure}[H]
    \centering
    \includegraphics[width=0.7\textwidth]{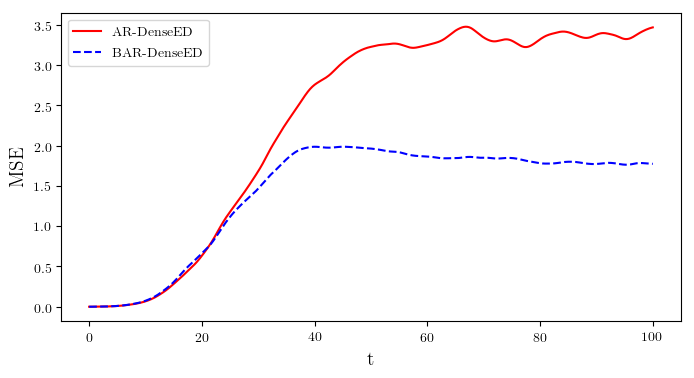}
    \caption{The mean MSE as a function of time for a test set of $200$ cases for the Kuramoto-Sivashinsky system.
    The error of BAR-DenseED is calculated using the expected value of the predictive distribution approximated using 30 samples of the posterior.}
    \label{fig:ks-mse}
\end{figure}

\begin{figure}[H]
    \centering
    \includegraphics[width=0.7\textwidth]{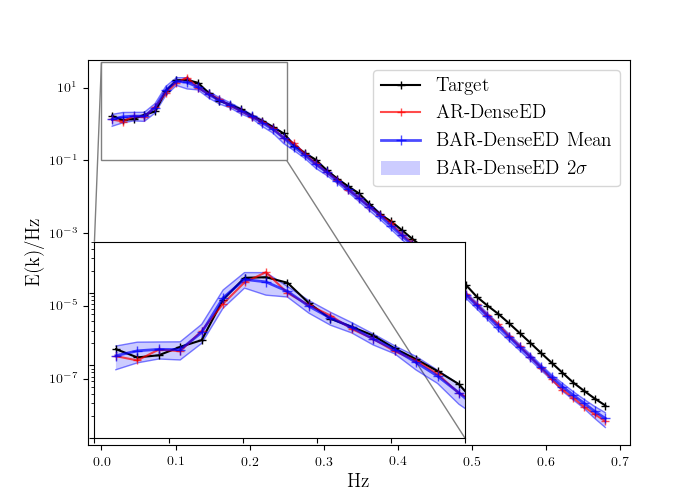}
    \caption{The time-averaged spectral energy density of the simulated result using the spectral ETDRK4 scheme (target), AR-DenseED deterministic prediction and BAR-DenseED empirical mean and standard deviation calculated from $30$ posterior samples.
    The averaged spectral energy density is the square of the modulus of the discrete Fourier transform over the domain, $x\in[0,22\pi]$, time-averaged between $t\in[0,500]$~\cite{brummitt2009search}.}
    \label{fig:ks-esd}
\end{figure}

For each test case, we calculate the spatial mean square error (MSE) at each time-step defined as:
\begin{equation}
    \textrm{MSE}(t) = \frac{1}{N}\sum^{N}_{i=1}\left(u_{i}(t)-u^{*}_{i}(t)\right)^{2},
    \label{eq:mse}
\end{equation}
where $N$, $u$ and $u^*$ are the total number of points used to discretize the domain, target value from the numerical simulator and the AR-DenseED prediction, respectively.
The mean of this error value is shown in Fig.~\ref{fig:ks-mse}, where we can see the decay in the model accuracy at around $t=20$ until AR-DenseED has fully diverged from the numerical simulator by $t>40$.
However, the predictions in Fig.~\ref{fig:ks-test-prediction} still appear to be physical despite not matching the numerical simulator.
To illustrate that the model predicts physical turbulence, we compare the average energy spectral density in Fig.~\ref{fig:ks-esd} for a randomly selected test case.
Since this is a turbulent statistic, the energy density profile is the same regardless of the particular initial condition.
This statistic was obtained by averaging time-steps between $t=[0,500]$.
This means our AR-DenseED is stable in its predictions for at least $5000$ time-steps, far beyond its training time-range.
This would not be possible with the traditional fully connected neural network approach for solving PDEs.
The AR-DenseED is accurate with the simulation results for the larger wavelengths where the majority of the energy is concentrated.
Additionally the AR-DenseED is able to correctly generate turbulence with the greatest energy at a similar wavelength as the simulation (Hz$\in[0.1,0.15]$).
While the model and simulation results begin to deviate from the numerical solution for smaller wavelengths, the energy decays at these higher frequencies meaning that the absolute error between the model and simulation is significantly less.
Thus we confidently conclude that the AR-DenseED has truly learned to predict physical turbulence of the K-S equation.

The average prediction wall-clock time of both the spectral ETDRK4 scheme versus the AR-DenseED are given in Table~\ref{tab:ks-wallclock}.
In this situation, the AR-DenseED fails to be a computational effective surrogate model compared to the highly-efficient spectral method which is able to take advantage of fast Fourier transform.
However, the K-S system was able to provide an excellent illustration of how AR-DenseED can model complex, non-linear, chaotic systems. We will show in the following sections how AR-DenseED can be computationally more efficient than traditional numerical solvers.
\begin{table}[H]
    \caption{Wall-clock time for both spectral ETDRK4 scheme and AR-DenseED to simulate $5000$ time-steps of the Kuramoto-Sivashinsky system.
    Wall-clock time estimates were obtained by averaging $10$ independent simulation run times.}
    \begin{tabular}{l|llcc}
               & \multicolumn{1}{c}{Hardware} & Backend & \multicolumn{1}{l}{$\Delta t$} & Wall-clock Time (s) \\ \hline
    Spectral   & Intel Xeon E5-2680           & Matlab  & $0.1$                           & \textbf{$0.185$}               \\
    AR-DenseED & Intel Xeon E5-2680           & PyTorch & $0.1$                           & $17.042$              \\
    AR-DenseED & GeForce GTX 1080 Ti          & PyTorch & $0.1$                           & $12.225$             
    \end{tabular}
    \label{tab:ks-wallclock}
\end{table}

\subsection{BAR-DenseED Probabilistic Predictions}
\noindent
To approximate the posterior with SWAG, $100$ samples of the model's parameters were collected.
This yielded reasonably diverse but accurate models and was found to be enough samples for $\bm{\theta}_{SWA}$ to converge.
During this period the learning rate was lowered to $1e-10$ for the neural network weights and $1e-6$ for the additive output noise.
While these learning rates may appear too small to sufficiently explore the local loss surface, for an auto-regressive model this was discovered to be a necessity as even very small changes to the parameters can have profound response changes during test time.
Larger learning rates for SWAG sampling were found to produce models that were unstable.

\begin{figure}[H]
    \centering
    \includegraphics[width=0.9\textwidth]{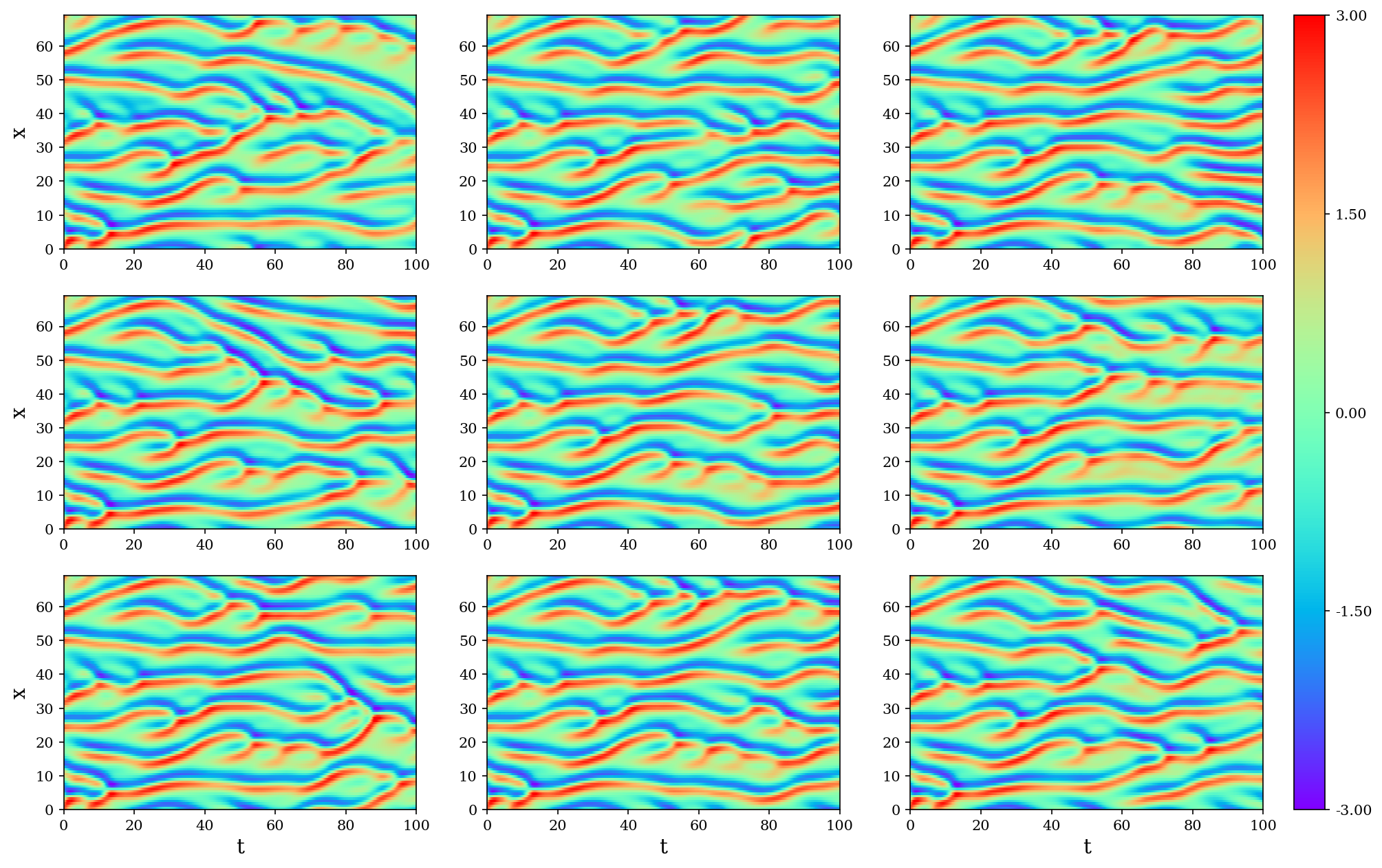}
    \caption{Samples from the posterior of BAR-DenseED approximated using SWAG for the Kuramoto-Sivashinsky system. The top left is the simulated result using the spectral ETDRK4 scheme.} 
    \label{fig:ks-bar-samples}
\end{figure}

We plot eight samples from the approximate posterior in Fig.~\ref{fig:ks-bar-samples} for a single test case with the target result in the top left.
Due to the chaotic nature of the K-S system or the so called ``butterfly effect'', samples from the posterior start with a similar response up for $t < 40$ and deviate for larger time values producing completely unique responses.
Similar to the deterministic case, we calculate the mean squared error defined in~\Eqref{eq:mse} using the expected predictive response using $30$ model samples for $200$ test cases and plot the mean error value at each time-step in Fig.~\ref{fig:ks-mse}.
Although, it appears BAR-DenseED performs much better than AR-DenseED for later time-steps this is due to the field being averaged out to around zero.
Thus the predictive performance between the deterministic and Bayesian model is essentially equivalent in this case.
We can propagate this uncertainty to the averaged spectral energy density illustrated in Fig.~\ref{fig:ks-esd} where $30$ model samples are used to calculate the spectral density.
At the largest wave lengths ($Hz < 0.3$), we can see that the model has reasonable error bars that are able to capture the true solution.
For smaller wave-lengths, the predicted energy density appears to be consistent between the samples.

% ==== 1D Viscous Burgers' Equation ====
\section{1D Viscous Burgers' Equation}
\label{sec:1dVisBurgers}
\begin{figure}[H]
    \centering
    \includegraphics[width=\textwidth]{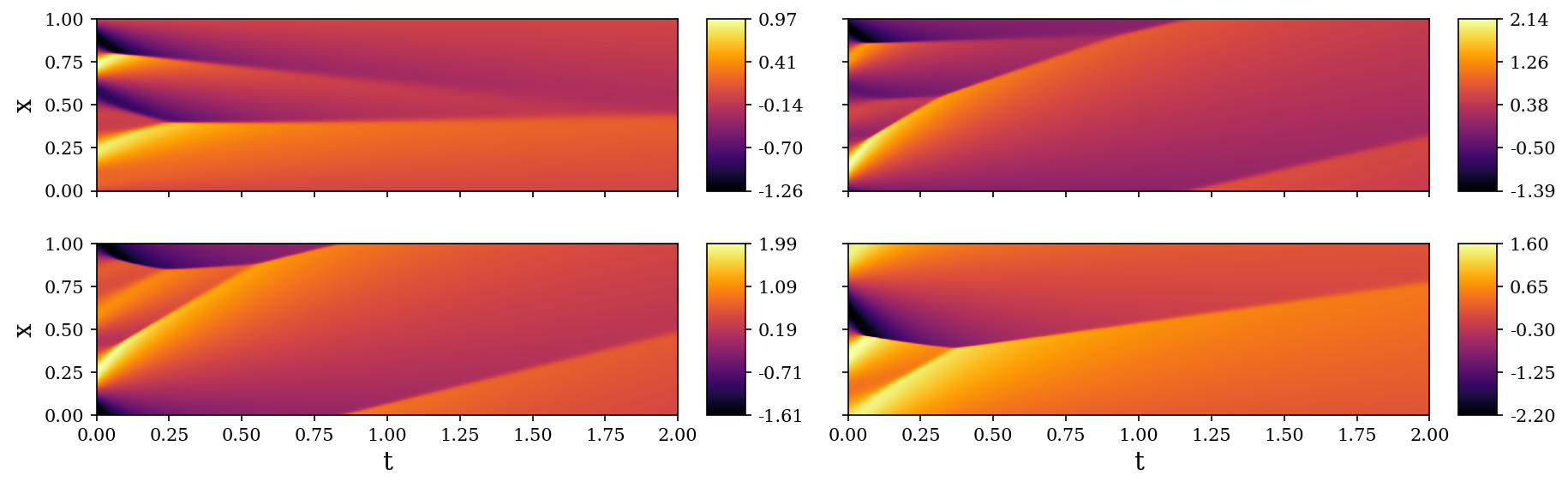}
    \caption{1D viscous Burgers' equation simulations for four various initial conditions solved using Fenics finite element package~\cite{alnaes2015fenics}.}
    \label{fig:burgers1D-FEM}
\end{figure}

\noindent
Let us now consider the 1D viscous Burgers' equation in a periodic domain:
\begin{gather}
    \frac{\partial u}{\partial t} + u\frac{\partial u}{\partial x} - \viscosity \frac{\partial^{2} u}{\partial x^{2}} = 0,\\
     u(0,t) = u(L,t), \quad x\in[0,L], \quad t\in[0,T],
\end{gather}
where $u$ is the velocity and $\viscosity$ is the viscosity.
The Burgers' equation is a fundamental PDE that arises in multiple areas ranging from fluid dynamics to traffic flow. It is most recognized for its characteristic shock formations~\cite{whitham2011linear}.
While cases of the 1D Burgers' equation have been recovered by machine learning models in the past~\cite{raissi2019physics}, ultimately these have been for relatively simple initial conditions consisting of a single shock.
Here, we would like to model much more complex dynamics by having a variable initial condition that contains multiple waves.
Consider a domain $x\in[0,1]$ with a constant viscosity of $\viscosity = 0.0025$ and the random initial condition given by a Fourier series with random coefficients:
\begin{equation}
    \begin{gathered}
    w(x) = a_0 + \sum_{l=1}^L a_l \sin(2l\pi x) + b_l \cos(2l\pi x), \\
    u(x,0) = \frac{2w(x)}{\max_x |w(x)|} + c,
    \label{eq:burger1d-initial}
    \end{gathered}
\end{equation}
where $a_l, b_l \sim \mc N(0, 1)$, $L=4$ and $c\sim \mc U(-1, 1)$.
In~\cite{pan2018long}, the authors   modeled a simpler random initial condition for the 1D viscous Burgers' system using a LSTM based model with some success.
However, this work reduced the complexity of the problem system by learning a reduced-order model rather than the true system.
Simulated system responses for several initial conditions are shown in Fig.~\ref{fig:burgers1D-FEM}.
We can see that the underlying dynamics of the system are fairly complex due to multiple shocks forming and then combining at later time-steps.
Shocks that intersect then combine and move in a different trajectory, making this a difficult system to predict accurately.
The auto-regressive model used for the 1D Burgers' system is similar to the one used for the K-S system in Section~\ref{sec:ks} with a few modifications.
For this system the five previous time-steps are used as inputs, $\bm{\chi}^{n+1}=\left\{\bm{u}^{n},\bm{u}^{n-1},\ldots,\bm{u}^{n-4}\right\}$ resulting in about $13000$ learnable weights.
The time-step value of the model is $\Delta t = 0.005$ with a spatial discretization of $512$ points.
This places the CFL number of the model well above one based on the maximum potential velocity for the specified random initial condition.
Again the negative log of the joint posterior in~\Eqref{eq:posterior-form} is the loss function with the implicit Crank-Nicolson time integration.
The remaining spatial gradients are discretized as follows:
\begin{gather}
    \begin{gathered}T_{\Delta t}(\bm{\mathcal{U}}^{n}, F_{\Delta x}) = \bm{u}^{n} + \Delta t \left[-0.5\left(F_{\Delta x}(\bm{x}, \bm{u}^{n+1}) + F_{\Delta x}(\bm{x}, \bm{u}^{n})\right)\right],\\
    F_{\Delta x}(\bm{x}, \bm{u}^{n}) = u^{n}u^{n}_{x} - \viscosity u^{n}_{xx},\end{gathered}\\
    uu_{x} = \frac{u^{2}_{i+1} - u^{2}_{i-1}}{4\Delta x}, \quad u_{xx} = \frac{u_{i+1} - 2u_{i} + u_{i-1}}{\Delta x^{2}},
\end{gather}
where the spatial gradients are approximated using second-order accurate approximations that are implemented efficiently using convolutional operators.
The model was trained for $100$ epochs with $2560$ training scenarios randomly sampled from~\Eqref{eq:burger1d-initial} and allowed to unroll a maximum of $200$ time-steps from its initial state.
Another $200$ samples from~\Eqref{eq:burger1d-initial} are used as a test set for assessing the models performance. 
Additional details on the model and training can be found in~\ref{app:burger1d}.

\subsection{AR-DenseED Deterministic Predictions}
\begin{figure}[H]
    \centering
    \includegraphics[width=\textwidth]{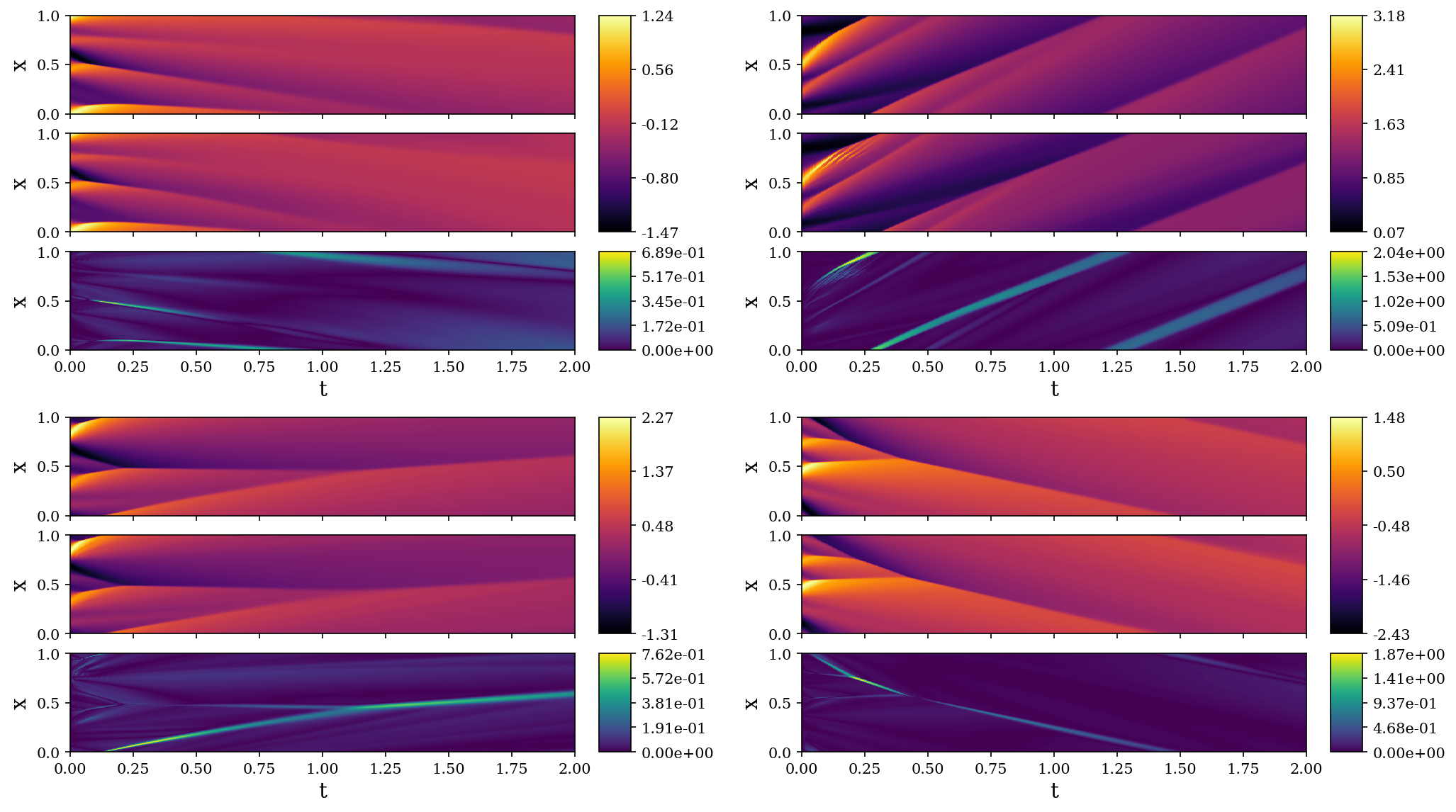}
    \caption{AR-DenseED predictions for four test initial conditions of the 1D Burgers' system. (Top to bottom) FEM target solution, AR-DenseED prediction, and $L_1$ error.}
    \label{fig:burgers1D-ARDenseED}
\end{figure}

\noindent
During testing we use the trained AR-DenseED to predict $400$ time-steps from its initial state.
This means that half of its prediction ($t>1.0$) is extrapolation beyond the time range used during training.
Four test cases are plotted in Fig.~\ref{fig:burgers1D-ARDenseED}, from which we can see that the AR-DenseED is able to predict this system accurately without any training data.
The target response is a high-fidelity finite element method (FEM) simulation at  time-step size $\Delta t = 0.001$, which is five times smaller than our surrogate model.
Overall, the AR-DenseED is able to accurately predict the shock formations and intersections with very distinct shock discontinuities.
In our past work~\cite{zhu2019physics}, we have found that convolutional neural networks like AR-DenseED can predict very sharp features much better than fully-connected models.
Additionally, we can see the reason why this system is difficult for surrogate modeling as one slight miscalculation in the shock intersection can result in compounding error.
This is illustrated in Fig.~\ref{fig:burgers1D-ARDenseED} where in some of the test cases the model gives a good prediction but a slight miscalculation in the shock intersection results in a growing error.
The excellent extrapolation capabilities of AR-DenseED are also shown for which the model is able to yield accurate predictions far beyond its initial training range.

\begin{figure}[H]
    \centering
    \includegraphics[width=0.9\textwidth]{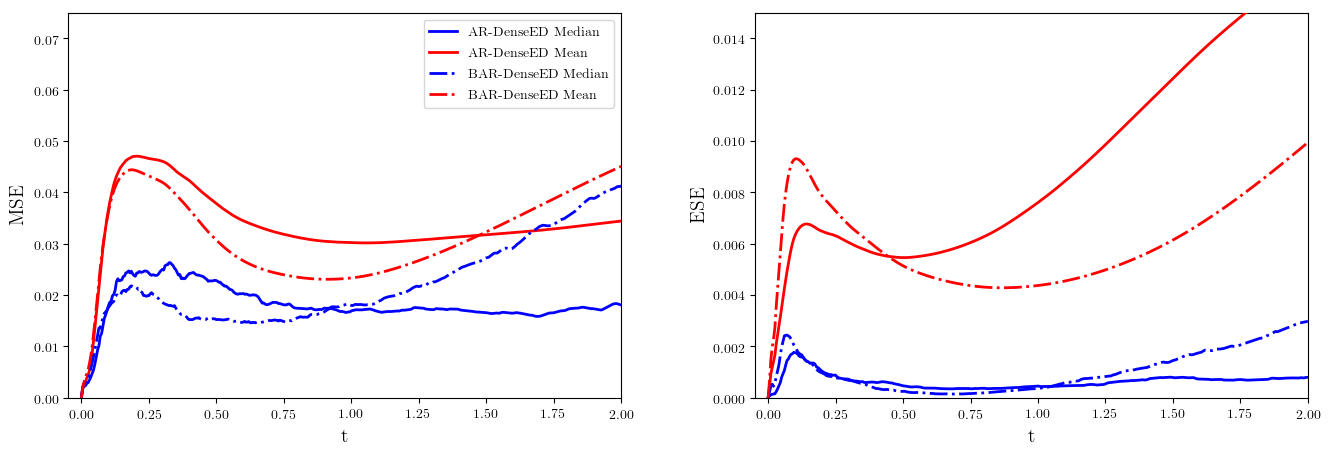}
    \caption{(Left to right) The mean square error (MSE) and energy squared error (ESE) as a function of time for a test set of $200$ cases for the 1D Burgers' system.
    The error of BAR-DenseED is calculated using the expected value of the predictive distribution approximating using $30$ samples of the posterior.}
    \label{fig:burgers1D-MSE}
\end{figure}

Now we consider the full $200$ test cases with target solutions provided by the high-fidelity FEM simulation.
For each test case, we calculate the spatial mean square error (MSE) defined in~\Eqref{eq:mse}.
In Fig.~\ref{fig:burgers1D-MSE}, we plot the mean and median of the MSE for the entire test set.
We can see an initial spike in the error during the initial shock formations/intersections that then decays as the system also decays.
However, the MSE can be slightly misleading to the actual quality of the prediction for this system since a small deviation in shock trajectory can potentially yield a growing error.
Thus, we also compute the energy square error (ESE) for a 1D domain:
\begin{align}
    \begin{split}
        \textrm{ESE}(t) &= \left[\int_{0}^{1} \frac{(u(x,t))^{2}}{2} dx - \int_{0}^{1}\frac{(u^{*}(x,t))^{2}}{2} dx\right]^{2},\\
        &= \left[\frac{1}{N}\sum^{N}_{i=1}\frac{(u_{i}(t))^{2}}{2} - \frac{1}{N}\sum^{N}_{i=1}\frac{(u_{i}^{*}(t))^{2}}{2}\right]^{2},
        \label{eq:ese}
    \end{split}
\end{align}
which instead captures the discrepancy of the total energy, $u^{2}/2$, within the domain, making this metric invariant to shock location.
Similarly, we plot the mean and median of the ESE for the $200$ test cases in Fig.~\ref{fig:burgers1D-MSE}.
For the ESE, we also see an initial spike followed by stable performance until the extrapolation region where the error begins to grow.
This behavior is to be expected as the model moves further from the initial training range.
Additionally, for both plots, we note that the mean error is consistently higher than the median which is a clear indication of outlier test cases that perform extremely poorly compared to the majority.
Unfortunately, this is a core drawback of the auto-regressive approach; if the initial prediction is poor this error will only grow as time progresses.
Finally, we compare the prediction computational cost of this surrogate model with both FEM and finite difference method (FDM) using fourth-order Runge-Kutta time integration in Table~\ref{tab:burger1d-wallclock}.
AR-DenseED is significantly computationally cheaper than the traditional methods, making it a potentially useful surrogate.
\begin{table}[H]
    \caption{Wall-clock time of finite element, finite difference and AR-DenseED to simulate $400$ time-steps of the 1D-Burgers' system.
    Wall-clock time estimates were obtained by averaging $10$ independent simulation run times.}
    \begin{tabular}{l|llcc}
                     & \multicolumn{1}{c}{Hardware} & Backend & \multicolumn{1}{c}{$\Delta t$} & Wall-clock (s) \\ \hline
    Finite Element    & Intel Xeon E5-2680  & Fenics  & $0.0005$  & $43.696$ \\
    Finite Element    & Intel Xeon E5-2680  & Fenics  & $0.001$  & $22.645$ \\
    Finite Element    & Intel Xeon E5-2680  & Fenics  & $0.005$   & $7.450$ \\
    Finite Difference & Intel Xeon E5-2680  & PyTorch & $0.0005$ & $4.856$ \\
    Finite Difference & GeForce GTX 1080 Ti  & PyTorch & $0.0005$ & $12.8359$ \\
    Finite Difference & Intel Xeon E5-2680  & PyTorch & $0.001$  & $2.862$ \\
    Finite Difference & GeForce GTX 1080 Ti   & PyTorch & $0.001$  & $6.264$\\
    AR-DenseED        & Intel Xeon E5-2680  & PyTorch & $0.005$  & $1.286$  \\
    AR-DenseED        & GeForce GTX 1080 Ti   & PyTorch & $0.005$  & $0.705$
    \end{tabular}
    \label{tab:burger1d-wallclock}
\end{table}
\subsection{BAR-DenseED Probabilistic Predictions}
\begin{figure}[H]
    \centering
    \includegraphics[width=0.9\textwidth]{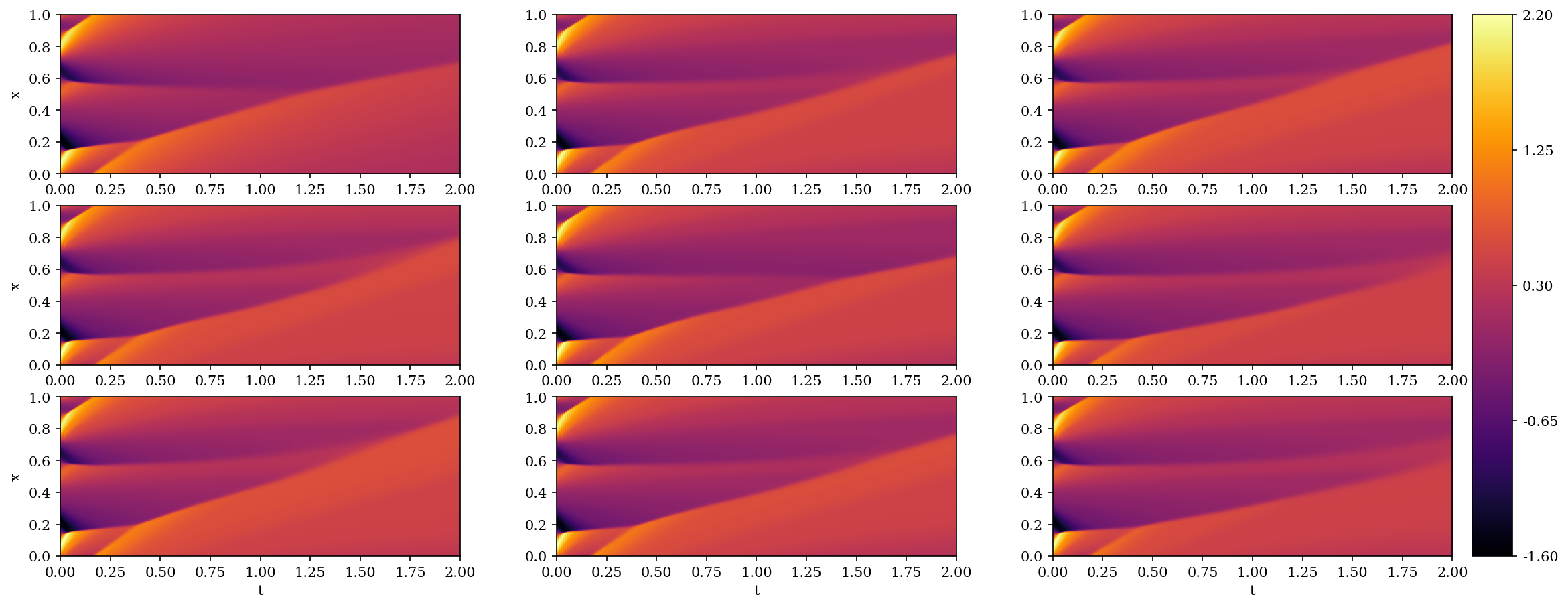}
    \caption{Samples from the posterior of BAR-DenseED approximated using SWAG for the 1D Burgers' system. The top left is the simulated result using the finite element method.} 
    \label{fig:burgers1D-bar-samples}
\end{figure}

\begin{figure}[H]
    \centering
    \includegraphics[width=\textwidth]{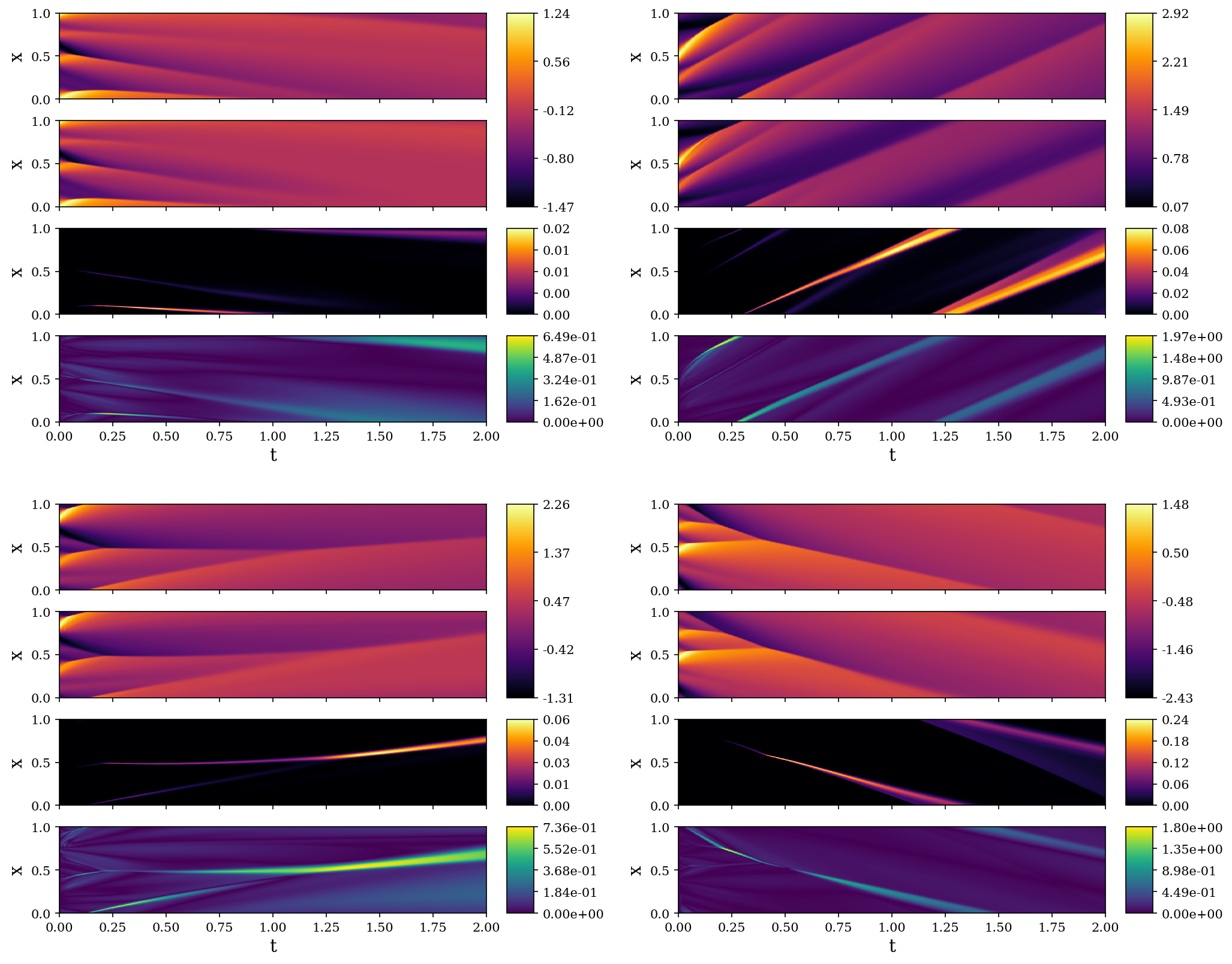}
    \caption{BAR-DenseED predictions for four test initial conditions of the 1D Burgers' system. (Top to bottom) FEM target solution, BAR-DenseED expected response, BAR-DenseED variance and $L_1$ error between the target and expected values.}
    \label{fig:burgers1D-BARDenseED}
\end{figure}

\noindent
For the posterior approximation, $90$ samples of the networks parameters were collected using a learning rate of $4e-8$.
Several samples from the posterior are illustrated in Fig.~\ref{fig:burgers1D-bar-samples} where slight differences in the predictions in earlier times change the final location of the wave at later times.
The predictive expectation and variance computed using $30$ model samples is plotted for four test cases in Fig.~\ref{fig:burgers1D-BARDenseED}.
We can clearly see that the bulk of the variance in the predictions occurs precisely where the shocks form/intersect as expected.
Similar to the deterministic model, we also plot the mean squared error and energy squared error of a test set of $200$ cases using the expected prediction of BAR-DenseED in Fig.~\ref{fig:burgers1D-MSE}.
The Bayesian framework is able to generally out perform the deterministic case.
One of the main reasons for this increase in accuracy is the reduction of outliers where AR-DenseED would yield an unsatisfactory prediction for only a small percentage of test cases.
Due to the model averaging in the predictive expectation, BAR-DenseED is able to more robustly handle these test cases where some individual predictions may be very poor.

\begin{figure}[H]
    \centering
    \includegraphics[width=\textwidth]{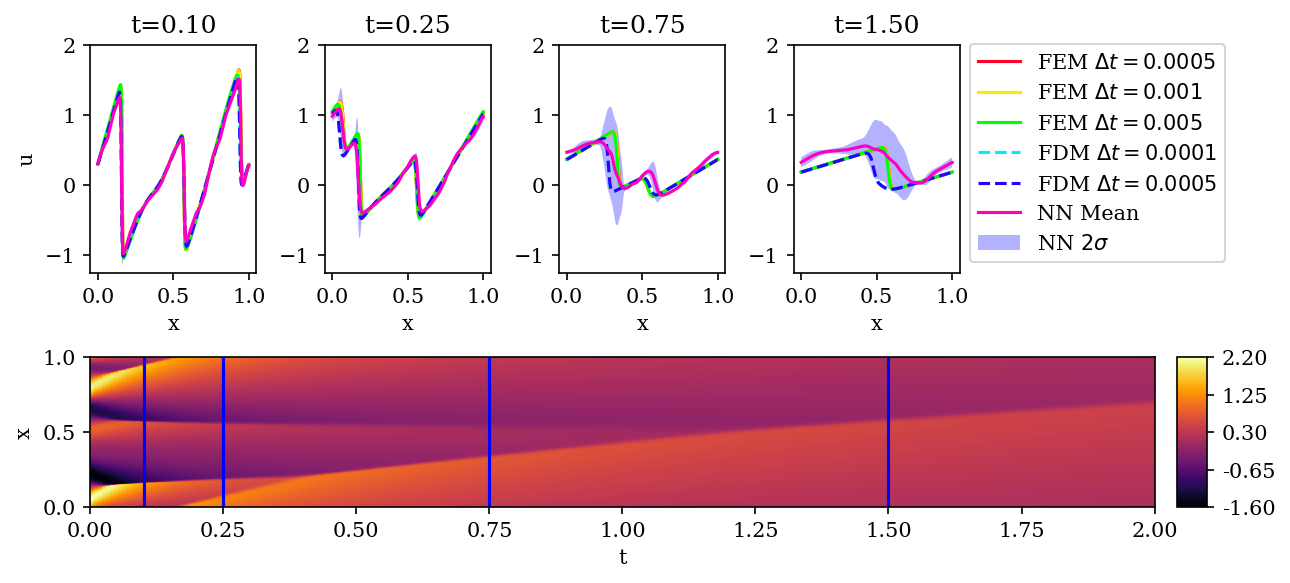}
    \caption{Instantaneous profiles of both the finite element method (FEM) and finite difference method (FDM) numerical solvers along with BAR-DenseED neural network (NN) predictive expectation and standard deviation at four various times of a test case.
    The bottom contour is the ideal target calculated using FEM with a time-step size $\Delta t =0.0005$.
    The blue lines mark each profile location.}
    \label{fig:burgers1D-profile1}
\end{figure}

\begin{figure}[H]
    \centering
    \includegraphics[width=\textwidth]{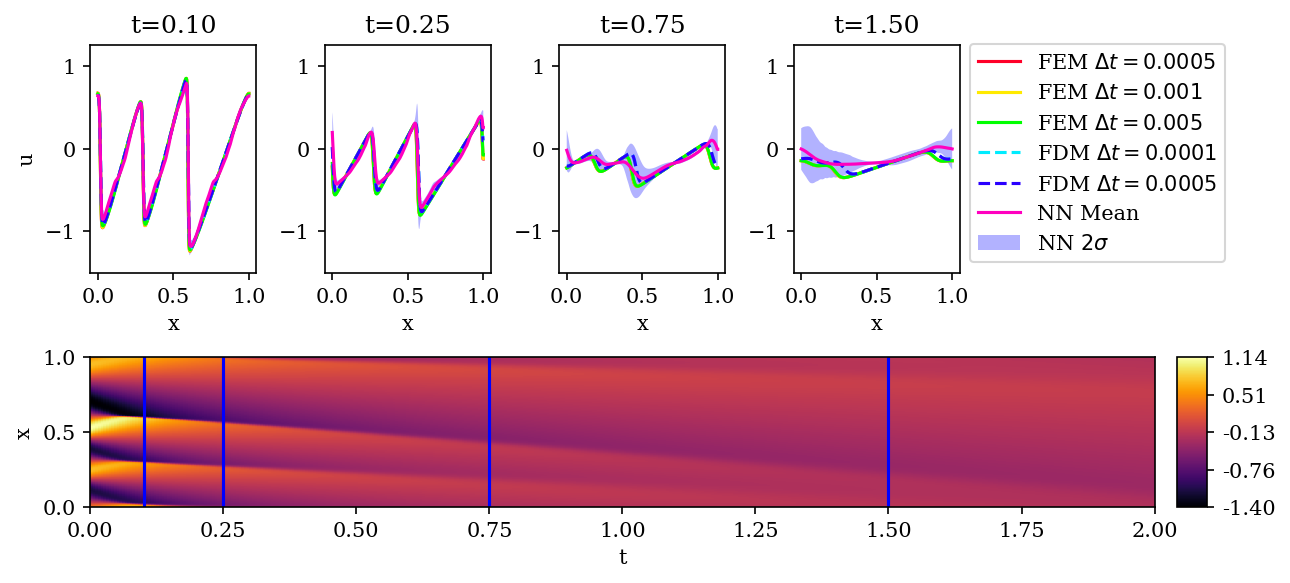}
    \caption{Instantaneous profiles of both finite element method (FEM) and finite difference method (FDM) numerical solvers along with BAR-DenseED neural network (NN) predictive expectation and standard deviation at four various times of a test case.
    The bottom contour is the ideal target calculated using FEM with a time-step size $\Delta t =0.0005$.
    The blue lines mark each profile location.}
    \label{fig:burgers1D-profile2}
\end{figure}

To better understand the uncertainty of BAR-DenseED and how it changes as the time series progresses, we plot several instantaneous profiles for two randomly selected test cases in Figs.~\ref{fig:burgers1D-profile1} and~\ref{fig:burgers1D-profile2}.
The first three profiles at $t=0.10$, $t=0.25$ and $t=0.75$ are within the time-range that was used during training.
The last profile at $t=1.50$ is considered extrapolation as it lies outside the training time-range.
In addition to the BAR-DenseED profile, the predicted solutions of the numerical solvers discussed in Table~\ref{tab:burger1d-wallclock} are also shown.
We note that there is a clear prediction discrepancy between the FEM and FDM solutions, largely due to  different numerical discretizations.
For this system, we hold the FEM solution as the higher accuracy method.

From both test cases, we can observe several important trends: the first is that for the earlier times the model compares well with the numerical solvers.
Second, as the shocks form, we can notice large spikes in the standard deviation at these locations which corresponds to the behavior seen in Fig.~\ref{fig:burgers1D-BARDenseED}.
Finally, we see the range of the error bars increases as the model begins to extrapolate indicating the model is less confident as it moves farther from its training range.
However, it is clear that in the extrapolated regions the model is still aware of the correct structure with the error bars capturing the true solution.

% ==== 2D Coupled Burgers' Equation ====
\section{2D Coupled Burgers' Equation}
\label{sec:2dVisBurgers}
\noindent
Lastly, we will consider the 2D coupled Burgers' system:
\begin{gather}
    \bm{u}_t + \bm{u}\cdot \nabla \bm{u} - \viscosity \Delta \bm{u} = 0,\\
    \bm{u}(0,y,t) = \bm{u}(L,y,t), \quad \bm{u}(x,0,t) = \bm{u}(x,L,t), \\
    \left\{x,y \right\} \in [0,L], \quad t\in[0,T],
\end{gather}
which when expanded into its components takes the following form:
\begin{equation}
    \begin{aligned}
        {\frac{\partial u}{\partial t}+u \frac{\partial u}{\partial x}+v \frac{\partial u}{\partial y} - \viscosity \left(\frac{\partial^{2} u}{\partial x^{2}}+\frac{\partial^{2} u}{\partial y^{2}}\right)} = 0, \\ 
        {\frac{\partial v}{\partial t}+u \frac{\partial v}{\partial x}+v \frac{\partial v}{\partial y} - \viscosity \left(\frac{\partial^{2} v}{\partial x^{2}}+\frac{\partial^{2} v}{\partial y^{2}}\right)} = 0,
    \end{aligned}
\end{equation}
where $\viscosity$ is the viscosity of the system which will be held at $\nu=0.005$ and the domain size set to $\left\{x,y \right\} \in [0,1]$.
$u$ and $v$ are the $x$ and $y$ velocity components, respectively.
The 2D coupled Burgers' equation is an excellent benchmark PDE due to both its non-linear term as well as diffusion operator, making it much more complex than the standard advection or diffusion equations.
The 2D coupled Burgers' belongs to a much broader class of PDEs that are related to various physical problems including shock wave propagation in viscous fluids, turbulence, super-sonic flows, acoustics, sedimentation and airfoil theory.
Given its similar form, the coupled Burgers' equation is often regarded as an essential stepping-stone to the full Navier-Stokes equations~\cite{ali2009computational, nee1998limit}.

\begin{figure}[H]
    \centering
    \includegraphics[width=0.9\textwidth]{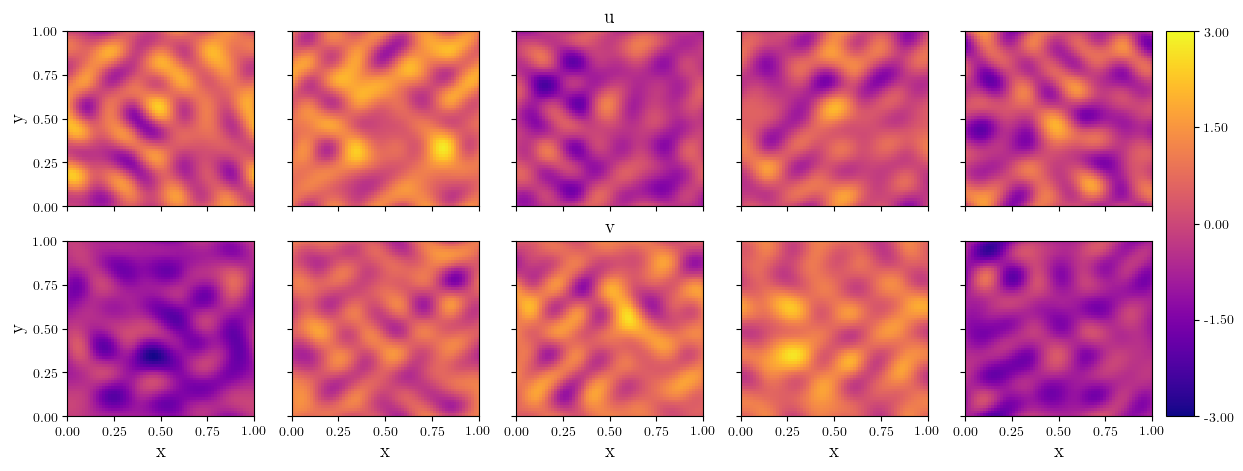}
    \caption{Randomly generated initial conditions for the 2D coupled Burgers' system. (Top to bottom) The $x$-velocity and $y$-velocity components. (Left to right) Different samples of the random initial condition.}
    \label{fig:burgers2D-Initial}
\end{figure}

\begin{figure}[H]
    \centering
    \begin{subfigure}{\textwidth}
        \includegraphics[width=\textwidth]{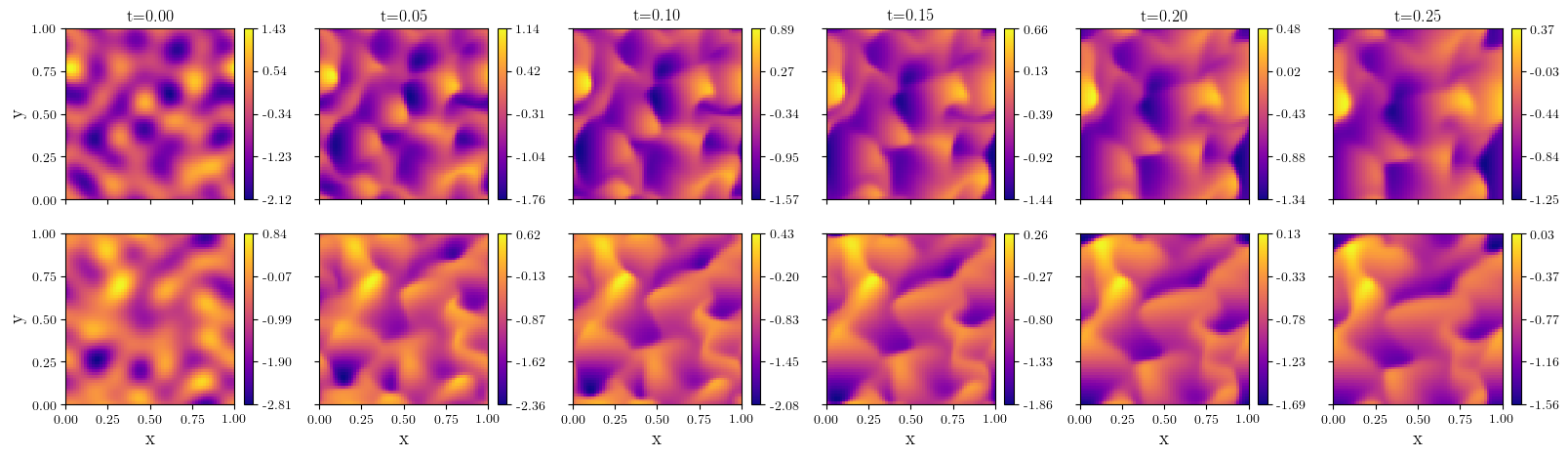}
        \vspace*{-7mm}
        \subcaption{Simulation 1}
    \end{subfigure}
    \begin{subfigure}{\textwidth}
        \includegraphics[width=\textwidth]{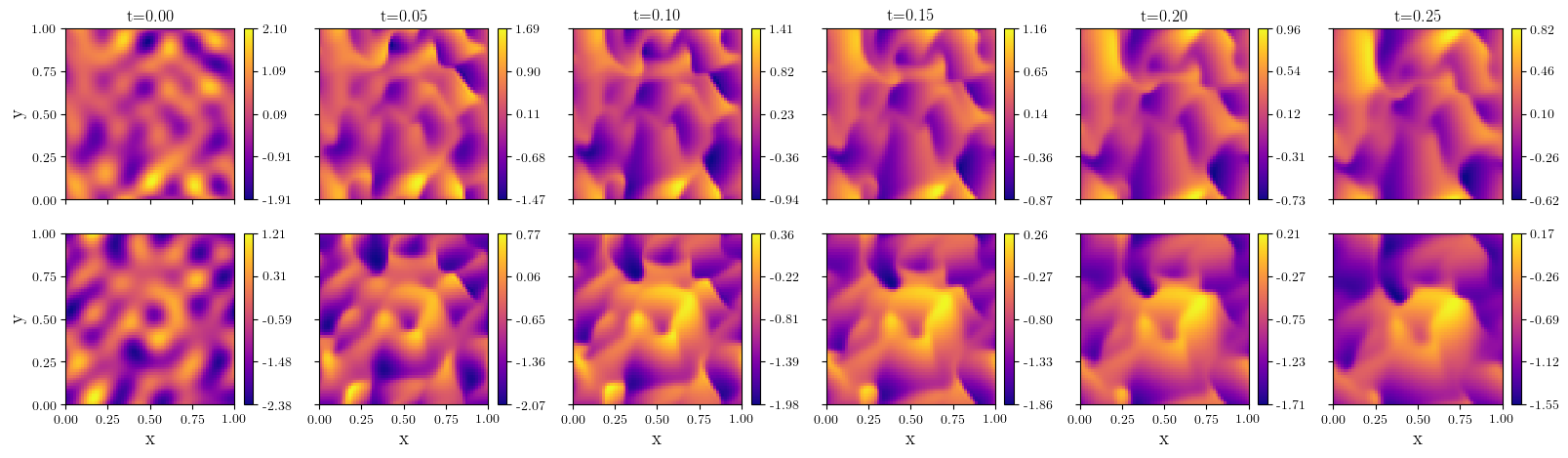}
        \vspace*{-7mm}
        \subcaption{Simulation 2}
    \end{subfigure}
    \caption{2D coupled Burgers' equation simulations for two various initial conditions solved using the Fenics finite element package~\cite{alnaes2015fenics}.
    (Top to bottom) $x$-velocity and $y$-velocity components.}
    \label{fig:burgers2D-fenics}
\end{figure}

As in our previous examples, we are interested in surrogate modeling for various initial conditions.
We will initialize the field using a truncated Fourier series with random coefficients:
\begin{equation}
    \begin{gathered}
        \bm{w}(x,y) = \sum_{i=-L}^L \sum_{j=-L}^L \bm{a}_{ij} \sin(2\pi\left(ix + jy\right)) + \bm{b}_{ij} \cos(2\pi\left(ix + jy\right)), \\
        \bm{u}(x, y, 0) = \frac{2\bm{w}(x,y)}{\max_{\left\{x,y\right\}} |\bm{w}(x,y)|} + \bm{c},
    \end{gathered}
    \label{eq:burger2d-initial}
\end{equation}
where $\bm{a}_{ij}, \bm{b}_{ij} \sim \mc N(0, \bm{I}_{2})$, $L=4$ and $\bm{c}\sim \mc U(-1, 1) \in \mathbb{R}^{2}$.
Several of these randomly generated initial conditions are illustrated in Fig.~\ref{fig:burgers2D-Initial}.
While these initial conditions may appear similar, the evolution of the systems results in the forming of many complex and unique structures.
To illustrate this, we plot several time-steps two FEM simulations of different initial conditions in Fig.~\ref{fig:burgers2D-fenics}.
As the system develops, we can see very distinct structures forming as waves form, mix, interact and dissipate with each other.
This is why the 2D coupled Burgers' system is a difficult system to model and serves as an excellent benchmark for our proposed surrogate.
The AR-DenseED model used for the 2D coupled Burgers' equations is the largest of the examples shown in this work with about $72000$ learnable parameters.
However, in the scope of the deep learning field over the past several years this network is still light weight.
Both the $x$ and $y$ velocity components are predicted by the same model in the form of two output channels.
Similarly both velocity components from three previous time-steps are used as inputs, $\bm{\chi}^{n+1}=\left\{\bm{u}^{n},\bm{v}^{n},\bm{u}^{n-1},\bm{v}^{n-1},\bm{u}^{n-2}, \bm{v}^{n-2}\right\}$, resulting in six input channels.
The time-step value of the model is $\Delta t = 0.005$ with a spatial discretization of $64 \times 64$ points.
Again the negative log of the joint posterior in~\Eqref{eq:posterior-form} is the objective function with the implicit Crank-Nicolson  time-integrator and other spatial gradients of the physics-constrained loss being discretized as:
\begin{gather}
    \begin{gathered}
    T_{\Delta t}(\bm{\mathcal{U}}^{n}, F_{\Delta x}) = \bm{u}^{n} + \Delta t \left[-0.5\left(F_{\Delta x}(\bm{x}, \bm{u}^{n+1}) + F_{\Delta x}(\bm{x}, \bm{u}^{n})\right)\right],\\
    F_{\Delta x}(\bm{x}, \bm{u}^{n}) = \bm{u}^{n}\cdot \nabla \bm{u}^{n} - \viscosity \Delta \bm{u}^{n},
    \end{gathered}\\
    \bm{u}_{x}=\frac{1}{8\Delta x}\begin{bmatrix} 
    -1 & 0 & 1 \\
    -2 & 0 & 2 \\
    -1 & 0 & 1
    \end{bmatrix}\ast \bm{u}, \quad 
    \bm{u}_{y}=\frac{1}{8\Delta x}\begin{bmatrix} 
        -1 & -2 & -1 \\
        0 & 0 & 0 \\
        1 & 2 & 1
    \end{bmatrix}\ast \bm{u}, \\
    \Delta \bm{u}=\frac{1}{2\Delta x^{2}}\begin{bmatrix} 
        1 & 0 & 1 \\
        0 & -4 & 0 \\
        1 & 0 & 1
    \end{bmatrix}\ast \bm{u},
\end{gather}
where the spatial gradients are approximated using Sobel filter 2D convolutions which are analogous to smoothed second-order accurate finite difference approximations~\cite{sobel19683x3}.
Using the Sobel filter was found to increase the stability of training and significantly reduce spurious oscillations in the model's predictions.
The model was trained on $5120$ training scenarios sampled from~\Eqref{eq:burger2d-initial} with a mini-batch size of $128$.
AR-DenseED was optimized for $100$ epochs and allowed to unroll a maximum of $100$ time-steps from its initial state.
Training on a single 1080Ti GPU required $9$ wall-clock hours.
To assess the performance of the model, another set of $200$ samples from~\Eqref{eq:burger2d-initial} was used as   testing scenarios.
Additional details on the model and training can be found in~\ref{app:burger2d}.

\subsection{AR-DenseED Deterministic Predictions}
\begin{figure}[H]
    \centering
    \includegraphics[width=\textwidth]{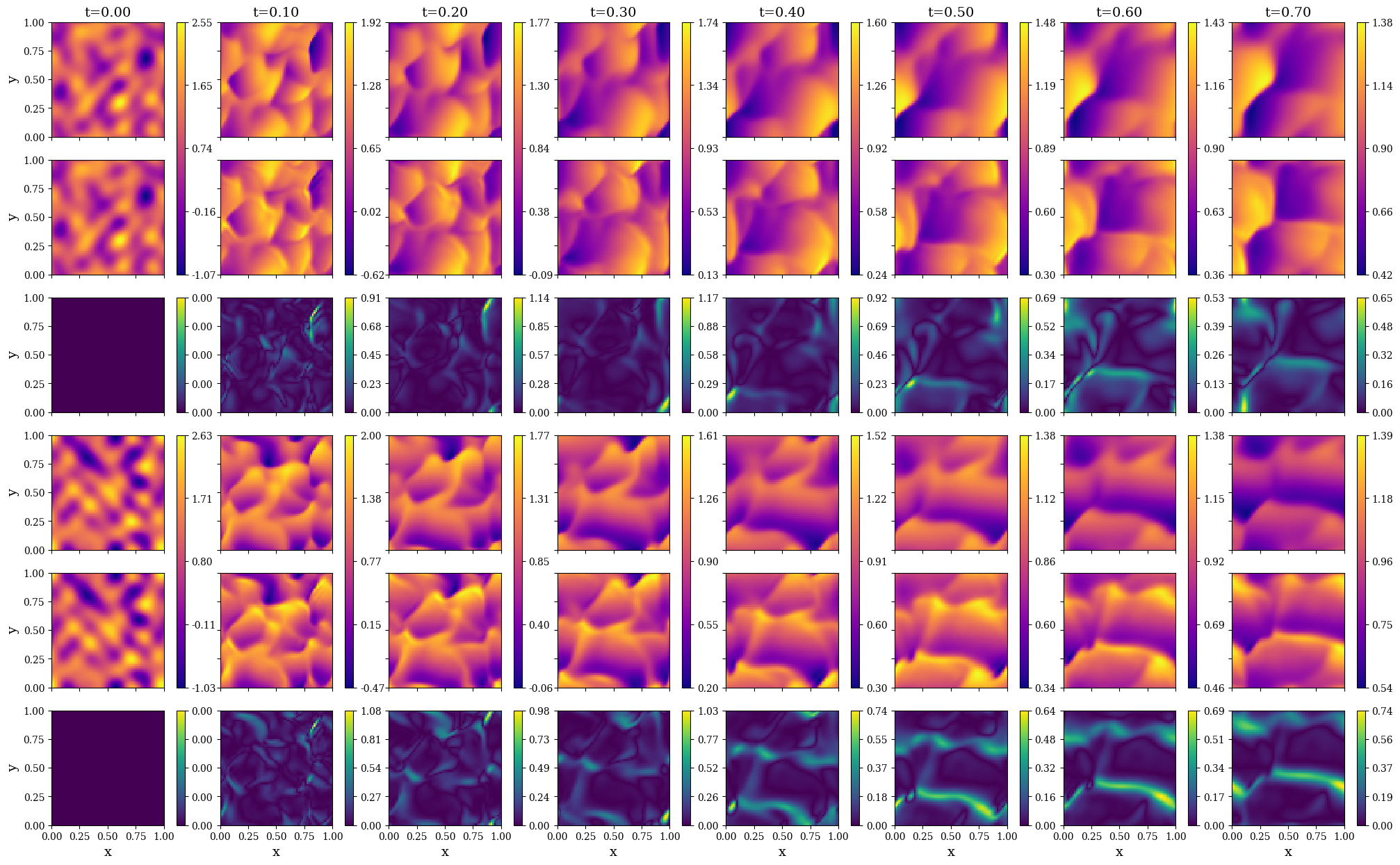}
    \caption{AR-DenseED predictions of a 2D coupled Burgers' test case. (Top to bottom) $x$-velocity FEM target solution, $x$-velocity AR-DenseED prediction, $x$-velocity $L_1$ error, $y$-velocity FEM target solution, $y$-velocity AR-DenseED prediction and $y$-velocity $L_1$ error.}
    \label{fig:burgers1D-ARDenseED-1}
\end{figure}

\begin{figure}[H]
    \centering
    \includegraphics[width=\textwidth]{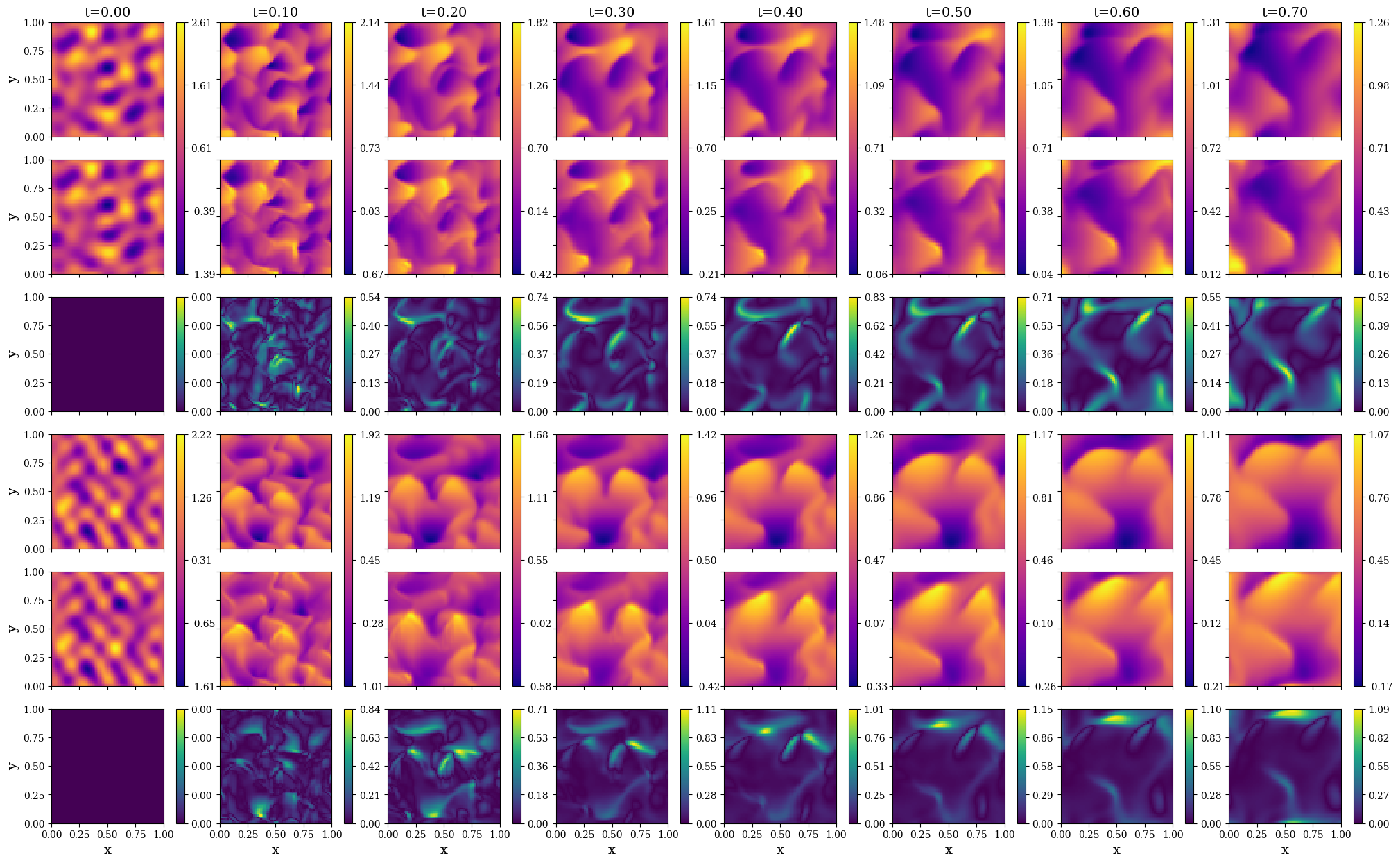}
    \caption{AR-DenseED predictions of a 2D coupled Burgers' test case. (Top to bottom) $x$-velocity FEM target solution, $x$-velocity AR-DenseED prediction, $x$-velocity $L_1$ error, $y$-velocity FEM target solution, $y$-velocity AR-DenseED prediction and $y$-velocity $L_1$ error.}
    \label{fig:burgers1D-ARDenseED-2}
\end{figure}

\noindent
For the 2D couple Burgers' system, the trained AR-DenseED is used to predict $200$ time-steps from the initial state at $t=0$.
Similar to the 1D Burgers' test case, this means that half of this predicted region ($t > 0.5$) is extrapolation beyond the time range that the model was trained on.
The target response is a high-fidelity FEM simulation with a discretization of $128 \times 128$ interpolated to a $64 \times 64$ grid.
The AR-DenseED's predictions are shown for two test cases in Figs.~\ref{fig:burgers1D-ARDenseED-1} and~\ref{fig:burgers1D-ARDenseED-2} in which several instantaneous time-steps are plotted.
Overall, the model does a remarkable job accurately predicting the complex structures and discontinuities of this system even into the extrapolation region.
As expected the bulk of the error is concentrated on shock interfaces and wave fronts, however,   the model's extremely good predictive capability is clear.

\begin{figure}[H]
    \centering
    \includegraphics[width=0.9\textwidth]{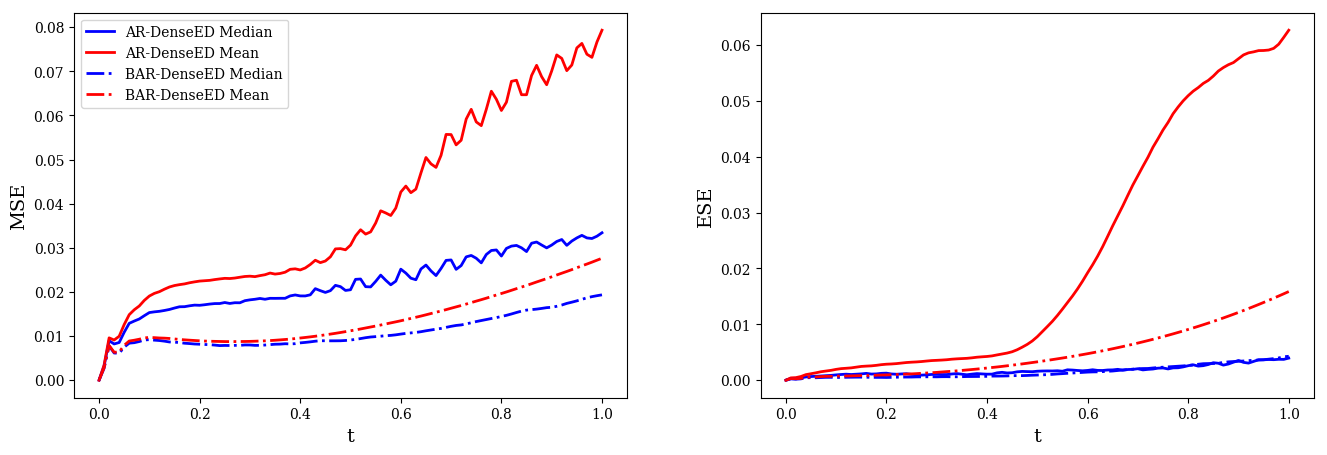}
    \caption{(Left to right) The mean square error (MSE) and energy squared error (ESE) as a function of time for a test set of $200$ cases for the 2D coupled Burgers' system.
    The error of BAR-DenseED is calculated using the expected value of the predictive distribution approximating using $30$ samples of the posterior.}
    \label{fig:burgers2D-MSE}
\end{figure}

The mean squared error defined in~\Eqref{eq:mse} as well as the energy squared error defined in~\Eqref{eq:ese}, both generalized to two dimensions, are evaluated for each time-step for the $200$ test scenarios.
The mean and the median of these error values for the entire test set are plotted in Fig.~\ref{fig:burgers2D-MSE}.
AR-DenseED is shown to have very stable prediction error within the training time region.
When AR-DenseED performs extrapolatory predictions at $t>0.5$, we can see that the mean error begins to grow faster than the median suggesting that several outlier test cases have very poor predictions.
We also compared the prediction computational cost of AR-DenseED for this 2D system again a set of FEM solutions of various spatial discretizations in Table~\ref{tab:burger2d-wallclock}.
Here, we can see that the AR-DenseED is able to far outperform the FEM simulations with a wall-clock time that is significantly less than all simulations by several orders of magnitude.
When one takes full advantages of a GPU hardware accelerator, this speed up becomes even larger.
Additionally this neglects the ability of the neural network to easily compute multiple simulations in a single batch, making its predictive efficiency even greater.
This illustrates the true potential power of these deep convolutional surrogate models.
\begin{table}[H]
    \caption{Wall-clock time of finite element simulation and AR-DenseED to simulate $200$ time-steps of the 2D coupled Burgers' system.
    Wall-clock time estimates were obtained by averaging $10$ independent simulation run times.}
    \begin{tabular}{l|llccc}
                     & \multicolumn{1}{c}{Hardware} & Backend & \multicolumn{1}{c}{$\Delta t$} & \multicolumn{1}{c}{$\Delta x$} & Wall-clock (s) \\ \hline
    Finite Element    & Intel Xeon E5-2680  & Fenics  & $0.005$  & 1/128 & $2955.38$ \\
    Finite Element    & Intel Xeon E5-2680  & Fenics  & $0.005$  & 1/64 & $418.83$ \\
    Finite Element    & Intel Xeon E5-2680  & Fenics  & $0.005$  & 1/32 & $133.65$ \\
    AR-DenseED        & Intel Xeon E5-2680  & PyTorch & $0.005$  & 1/64 & $4.691$  \\
    AR-DenseED        & GeForce GTX 1080 Ti   & PyTorch & $0.005$  & 1/64 & $0.841$
    \end{tabular}
    \label{tab:burger2d-wallclock}
\end{table}

\subsection{BAR-DenseED Probabilistic Predictions}
\begin{figure}[H]
    \centering
    \includegraphics[trim={0 0 0 1cm}, clip, width=0.9\textwidth]{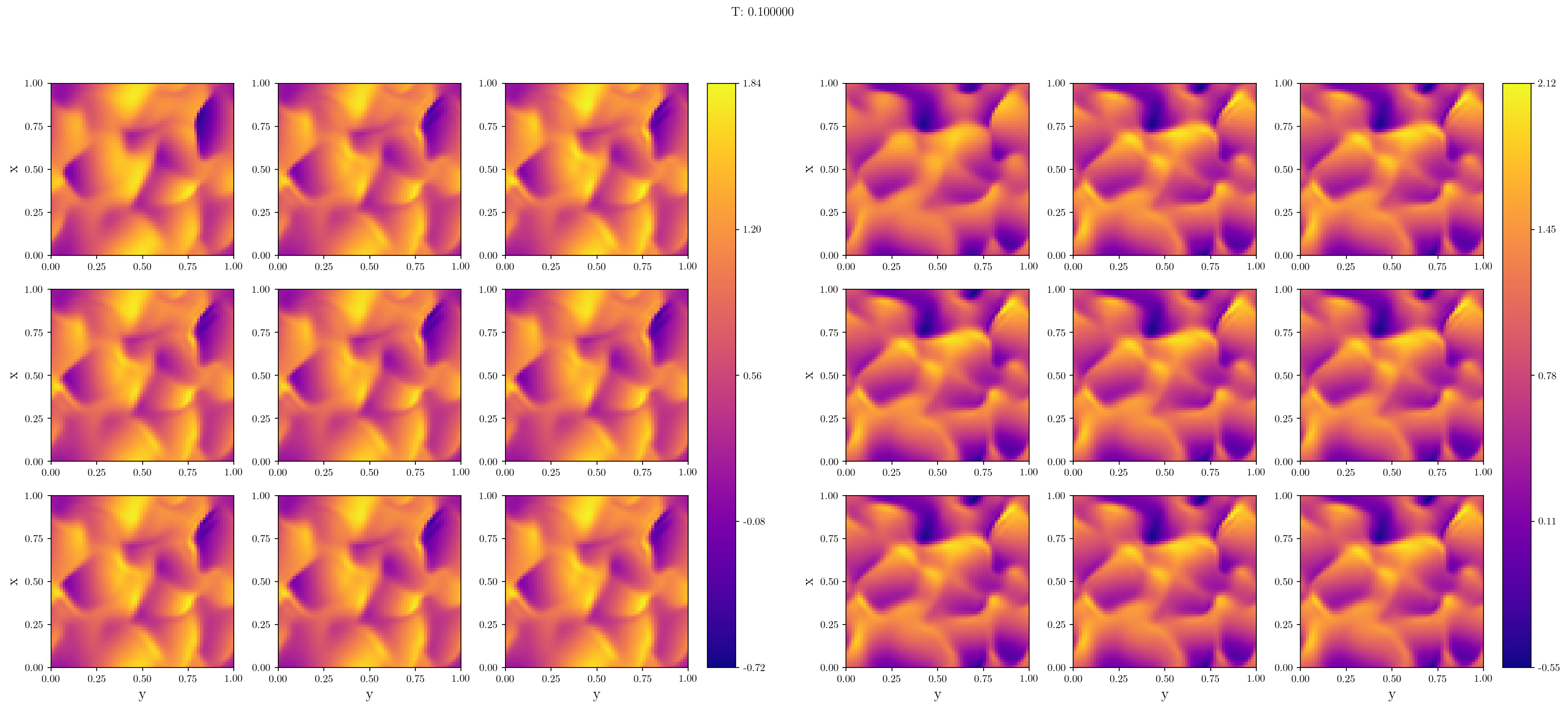}
    \caption{(Left to right) Samples of the $x$-velocity and $y$-velocity component from the posterior of BAR-DenseED approximated using SWAG at $t=0.1$ for the 2D coupled Burgers' system. The top left in each grid is the simulated result using FEM.} 
    \label{fig:burgers2D-bar-samples-1}
\end{figure}

\begin{figure}[H]
    \centering
    \includegraphics[trim={0 0 0 1cm}, clip, width=0.9\textwidth]{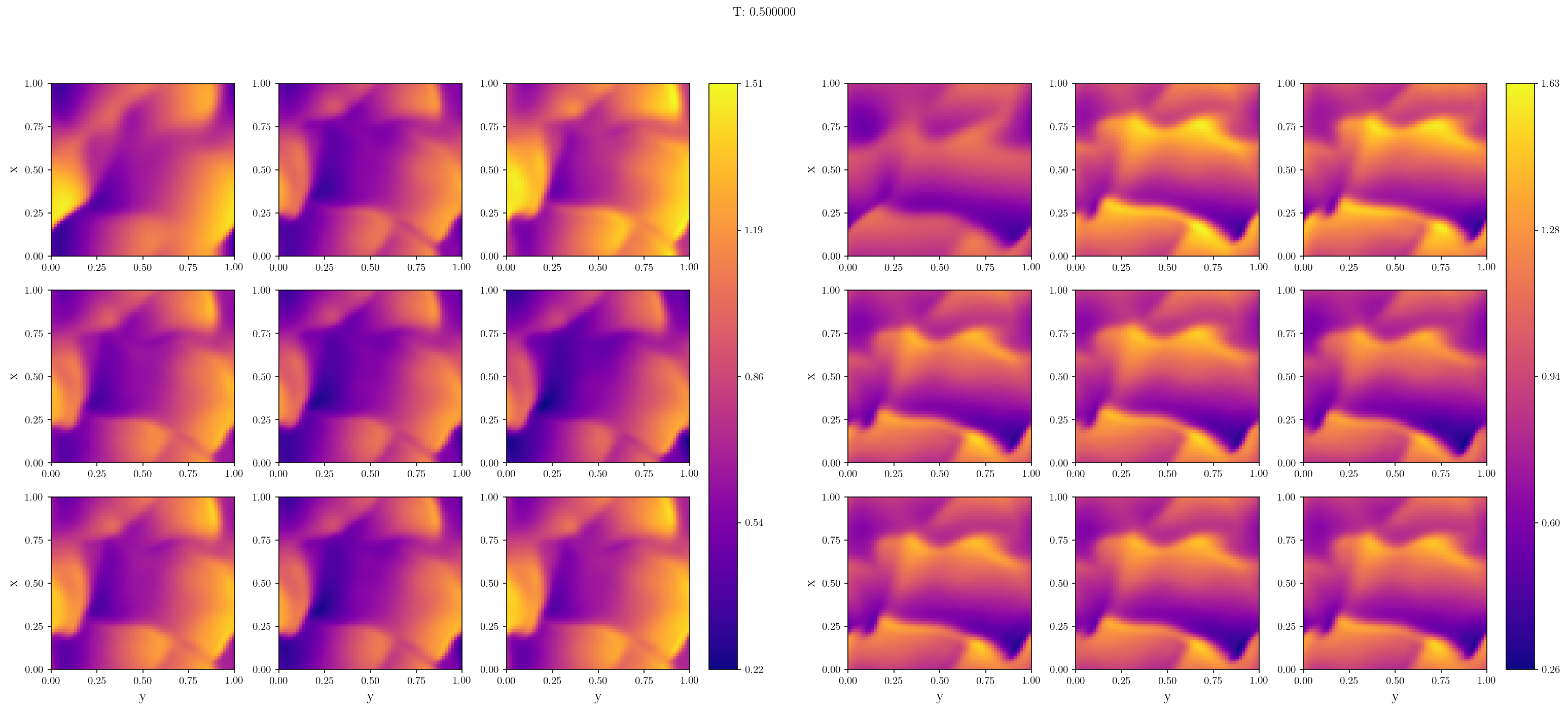}
    \caption{(Left to right) Samples of the $x$-velocity and $y$-velocity component from the posterior of BAR-DenseED approximated using SWAG at $t=0.5$ for the 2D coupled Burgers' system. The top left in each grid is the simulated result using FEM.} 
    \label{fig:burgers2D-bar-samples-2}
\end{figure}

\begin{figure}[H]
    \centering
    \includegraphics[width=\textwidth]{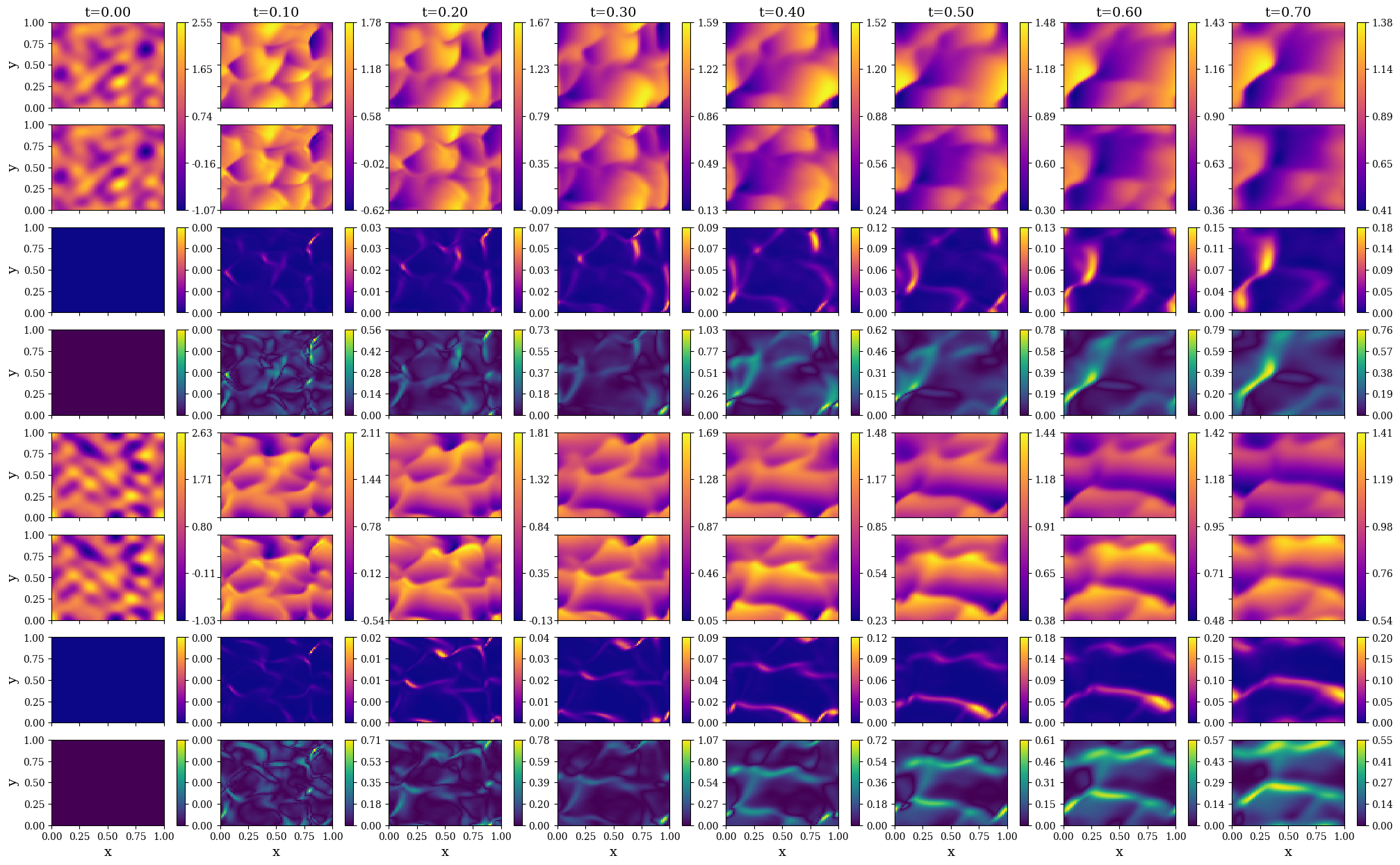}
    \caption{BAR-DenseED predictions for a 2D coupled Burgers' test case. (Top to bottom) $x$-velocity FEM target solution, BAR-DenseED expected response, BAR-DenseED variance, $L_1$ error between the target and expected values and similarly followed by the $y$-velocity component. }
    \label{fig:burgers2D-BARDenseED-1}
\end{figure}

\begin{figure}[H]
    \centering
    \includegraphics[width=\textwidth]{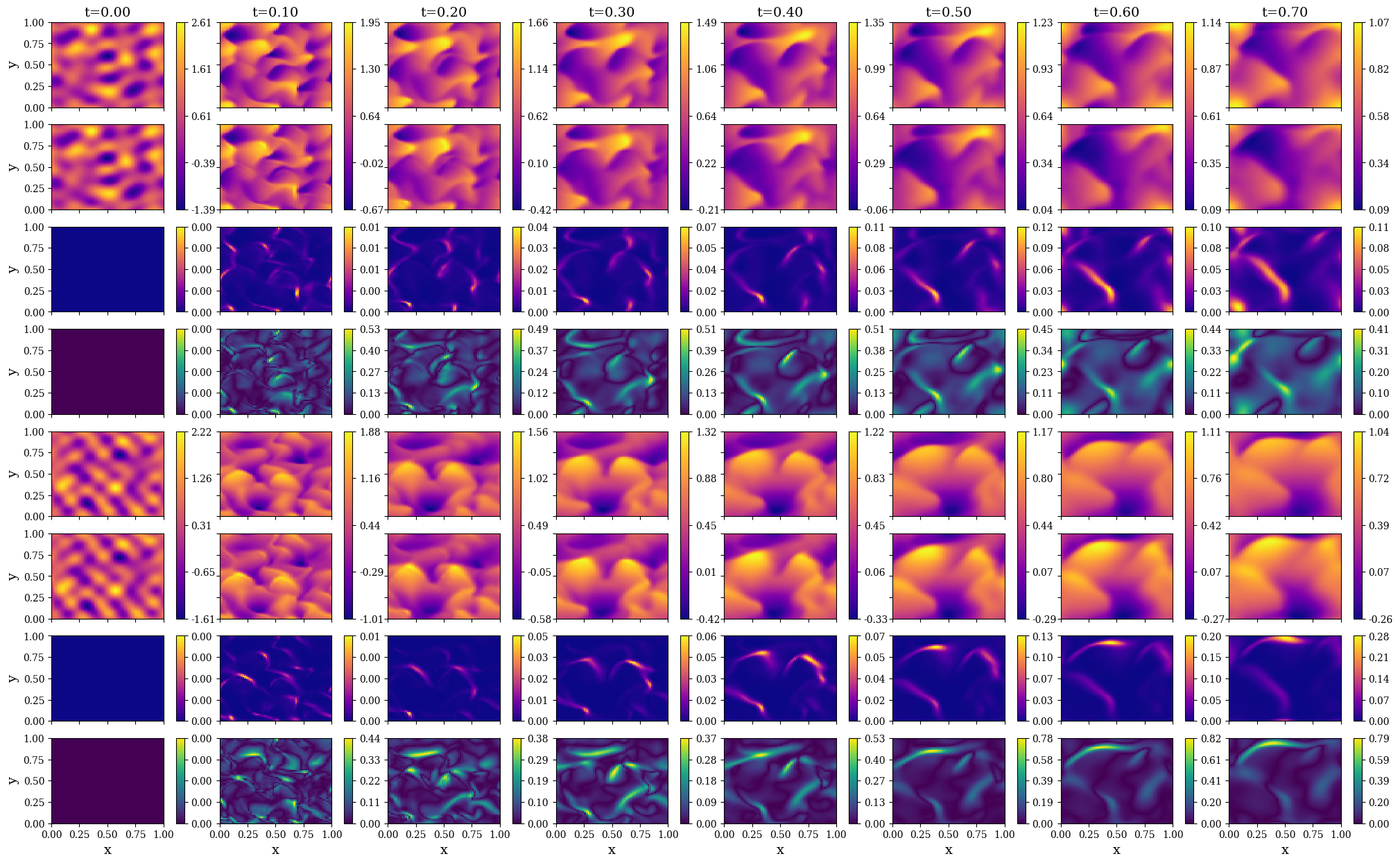}
    \caption{BAR-DenseED predictions for a 2D coupled Burgers' test case. (Top to bottom) $x$-velocity FEM target solution, BAR-DenseED expected response, BAR-DenseED variance, $L_1$ error between the target and expected values and similarly followed by the $y$-velocity component.}
    \label{fig:burgers2D-BARDenseED-2}
\end{figure}

\begin{figure}[H]
    \centering
    \includegraphics[width=\textwidth]{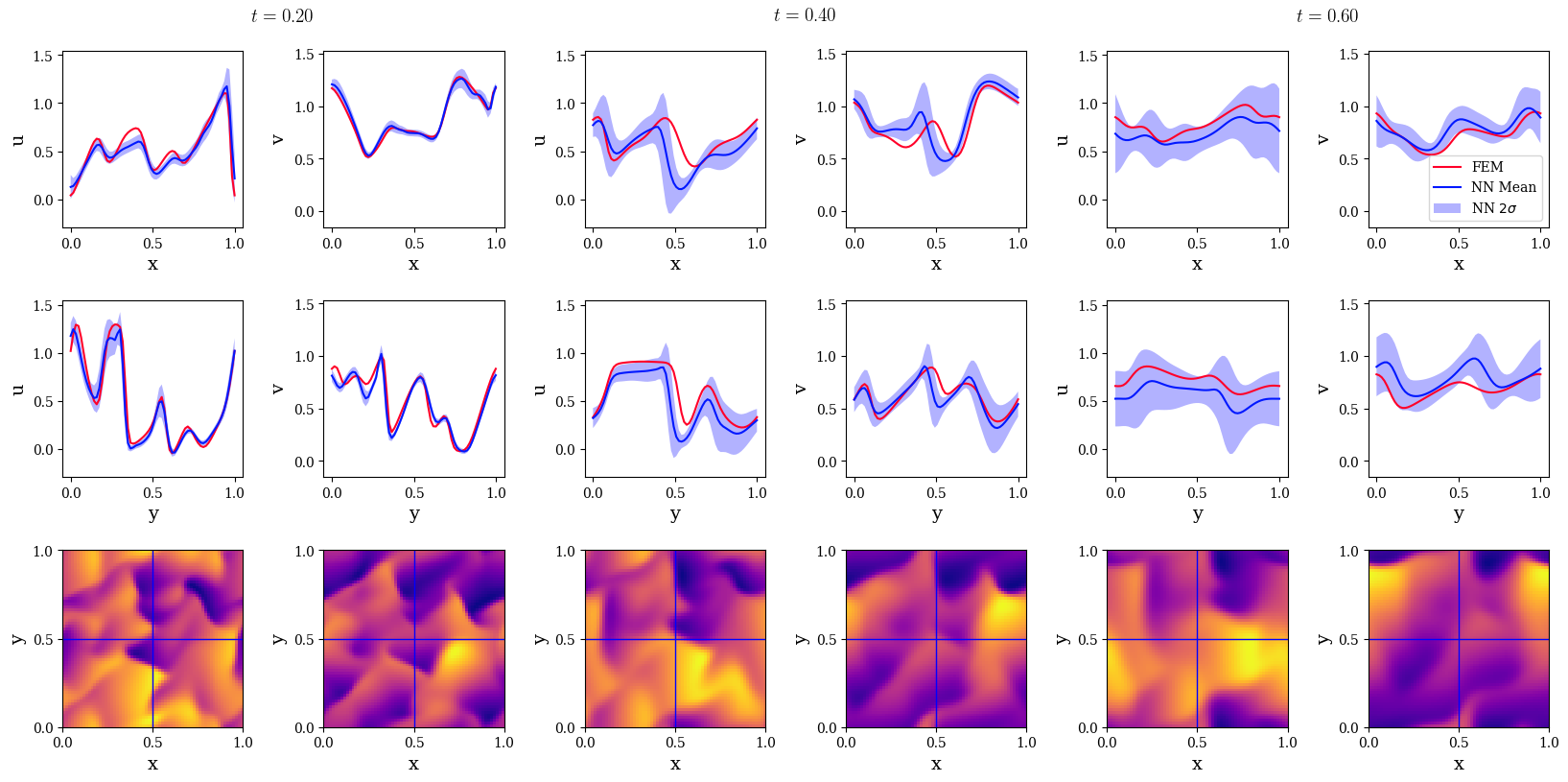}
    \caption{Instantaneous profiles of the finite element method (FEM) solver and BAR-DenseED (NN) predictive expectation and standard deviation at three various times of a test case.
    (Top to bottom) Horizontal profile at $y=0.5$, vertical profile at $x=0.5$ and target FEM contour with blue lines to show the profile locations.
    (Left to right) $x$-velocity and $y$-velocity profiles at $t=0.2$, $0.4$ and $0.6$.}
    \label{fig:burgers2D-profiles}
\end{figure}

\noindent
To approximate the posterior with SWAG, $90$ samples of the network's parameters were collected with a learning rate of $3e-8$ for the neural network parameters and $3e-5$ for the output noise $\beta$.
Predictions of several posterior samples at times $t=0.1$ and $t=0.5$ are shown in Figs.~\ref{fig:burgers2D-bar-samples-1} and~\ref{fig:burgers2D-bar-samples-2}, respectively.
As expected, the variance of the samples increases significantly as time progresses.
This is also reflected in the BAR-DenseED prediction contours for two test cases in Figs.~\ref{fig:burgers2D-BARDenseED-1} and~\ref{fig:burgers2D-BARDenseED-2} in which the predictive expectation and variance computed using $30$ model samples are shown for several time-steps.
Again we see that the majority of the uncertainty is concentrated on the leading face of the shocks/waves which is precisely where we would expect it for a well calibrated model.
To more clearly illustrate the uncertainty estimates of the probabilistic model, velocity profiles of both velocity components are plotted in Fig.~\ref{fig:burgers2D-profiles} for a randomly selected test case.
Overall, we can see that the predictive standard deviation is able to capture the true solution for almost all times.
Finally, the mean squared error and energy squared error are also plotted using the predictive expectation of BAR-DenseED in Fig.~\ref{fig:burgers2D-MSE}.
The Bayesian model is able to have smaller and more stable prediction error even when extrapolating beyond the training time range.

% ==== Conclusion ====
\section{Conclusion}
\label{sec:conclusion}
\noindent
In this work, we have presented a deep auto-regressive convolutional neural network model that can be used to learn and surrogate model the dynamics of transient PDEs.
To train this model, physics-constrained deep learning is used where the governing equations of the system of interest are used to formulate a loss function.
This allows the model to be trained with zero (output) training data.
Additionally, we proposed a Bayesian probabilistic framework built on top of this deep learning model to allow for uncertainty quantification (including both epistemic and aleatoric uncertainty).
This model was implemented for three PDE systems: the first is the chaotic Kuramoto-Sivashinsky equation for which the model was used to accurately reproduce physical turbulent statistics.
The second is the 1D Burgers' equation at a low viscosity where the model was able to successfully predict multi-shock wave formation and intersections.
At last is the 2D coupled Burgers' equations for which the model was able to accurately predict the complex wave dynamics of this system.
Overall, the proposed model showed exceptional predictive accuracy and was able to successfully extrapolate to predict outside the time-range used when training.

Although fully connected networks are frequently used to solve PDE systems due to their analytical and mesh-less benefits, the performance of convolutional neural networks for solving and surrogate modeling of PDEs is exceptional.
In this work, we have further shown that convolutional neural networks can be used effectively in physics-constrained learning and build surrogate models that are order of magnitudes faster than state-of-the-art numerical solvers.
A particular draw-back of convolutional neural networks is the requirement that both spatial and temporal derivatives be discretized, which opens the model up to the challenges that are faced in traditional numerical algorithms such as truncation error, oscillations, convergence criterion and more.
However, one can also use the deep repository of techniques and tricks developed by the numerical analysis community to address these potential issues.
This would be an interesting avenue to investigate as one could incorporate methods such as flux limiters or non-oscillatory schemes to yield predictions that have similar numerical benefits.
One could also consider higher-order derivatives and their impact on accuracy versus training stability.

The most obvious path to further develop this model is to implement it for more complex and larger systems.
This could include systems such as the Navier-Stokes equations, coupled transport through porous media, combustion and more.
However, there are still significant challenges that will need to be addressed.
The most important is training cost; with any time series problem training a deep learning model becomes exponentially more difficult and more costly.
Although our model is able to be trained in a very reasonable amount of time given the complexity of the physical systems modeled as well as the hardware used, improving the training of the model will still be an important area of study. 
This may involve the use of network architectures considered in recent neural language processing literature such as self-attention mechanisms.
Another potential extension is the incorporation of data and physics-constrained learning to create this hybrid learning framework.
Specifically for time series, one may not have the system state at every time interval that is desired.
Physics-constrained learning could be an answer to help bridge this challenge of predicting at fine resolutions with sparse data.

\section*{Acknowledgements}
\noindent
The authors acknowledge support  from the Defense Advanced Research Projects Agency (DARPA) under the Physics of Artificial Intelligence (PAI) program (contract HR$00111890034$). 
The work of NG is also supported by the National Science Foundation (NSF) Graduate Research Fellowship Program grant No. DGE-$1313583$. 
Additional computing resources were provided by the University of Notre Dame's Center for Research Computing (CRC), NSF supported ``Extreme Science and Engineering Discovery Environment'' (XSEDE) on the Bridges and Bridges-GPU cluster through research allocation No. TG-CTS$180038$ and by the AFOSR Office of Scientific Research through the DURIP program.

\appendix
% ======== Kuramoto-Sivashinsky Appendix ========
\section{Kuramoto-Sivashinsky}
\label{app:ks}
\begin{figure}[H]
    \centering
    \includegraphics[width=0.5\textwidth]{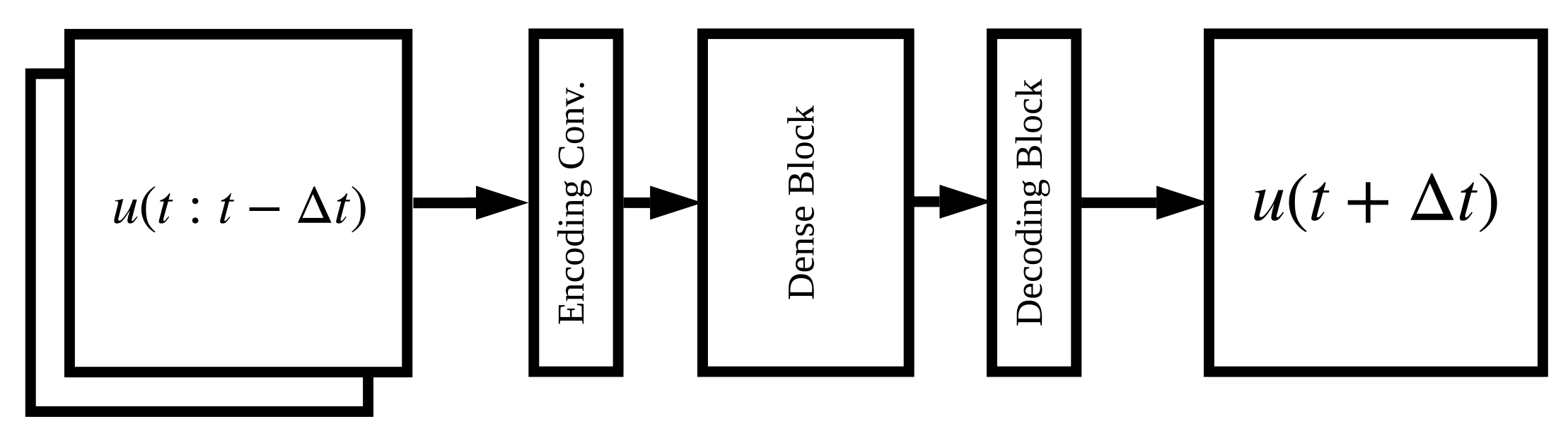}
    \caption{The AR-DenseED model with $4821$ learnable parameters used for the Kuramoto-Sivashinsky equation.
    This model consists of an encoding convolution, single dense block with a growth rate of $4$ and a length of $4$ followed by a decoding block.
    The two previous time-steps are used as inputs.}
    \label{fig:app-ks-model}
\end{figure}

\begin{figure}[H]
    \centering
    \includegraphics[width=0.9\textwidth]{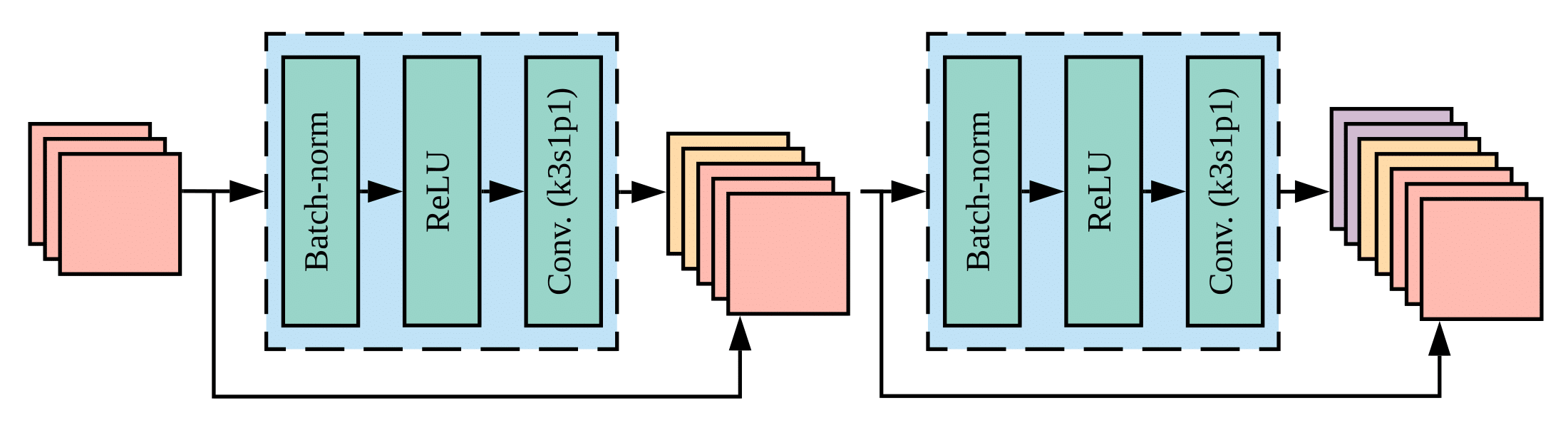}
    \caption{Schematic of a dense block with a growth rate of $2$ and length of $2$ consisting of batch-normalization~\cite{ioffe2015batch}, Rectified Linear Unit (ReLU) activation functions~\cite{glorot2011deep} and convolutions.
    The key feature is the residual connection that stacks the output of each convolution increasing the number of feature channels substantially.
    Convolutions are denoted by the kernel size $k$, stride $s$ and padding $p$.}
    \label{fig:app-ks-denseBlock}
\end{figure}

\begin{figure}[H]
    \centering
    \includegraphics[width=\textwidth]{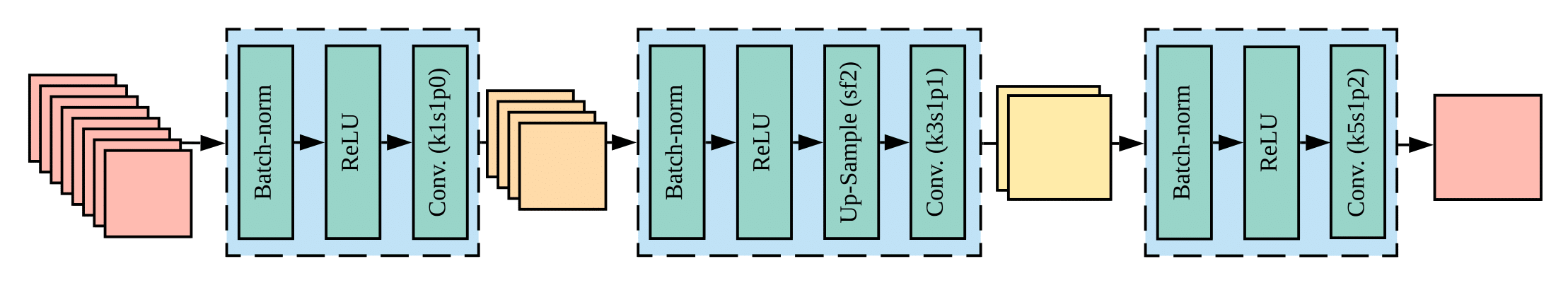}
    \caption{Schematic of a decoding block which consists of several sequential layers of batch-normalization~\cite{ioffe2015batch}, Rectified Linear Unit (ReLU) activation function~\cite{glorot2011deep} and convolutions.
    Nearest neighbor up-sampling is used to increase the size of the data to the desired dimensions.
    Convolutions are denoted by the kernel size $k$, stride $s$ and padding $p$.}
    \label{fig:app-ks-decodingBlock}
\end{figure}

\noindent
The following appendix discusses details related to the model used to predict in the Kuramoto-Sivashinsky equation in Section~\ref{sec:ks}.
For this system, a small dense encoder-decoder model was used as depicted in Fig.~\ref{fig:app-ks-model}.
The three components of this model are the encoding convolution, the dense block  and decoding block.
The resulting model encodes a given 1D input $\left\{\bm{u}^{n}, \bm{u}^{n-1}\right\}\in\mathbb{R}^{d}$ to a set of latent variables that are of dimensionality $\bm{z}_{i}\in\mathbb{R}^{d/2}$.
These latent variables are then decoded to the prediction $\bm{u}^{n+1}\in\mathbb{R}^{d}$.
Examples of a dense block and decoding block are shown in Fig.~\ref{fig:app-ks-denseBlock} and Fig.~\ref{fig:app-ks-decodingBlock} originally proposed in Zhu and Zabaras~\cite{zhu2018bayesian}.

While this model is relatively small, we found that smaller models were more stable in training and had the additional benefit of being faster.
To help with learning periodic boundaries, circular padding was used for all convolutions within the model.
The model was optimized with ADAM~\cite{kingma2014adam} with an initial learning rate of $1e-3$.
It was found that decaying the learning rate exponentially yielded the most stable, consistent and accurate results.
Additional model training parameters can be reference in Table~\ref{tab:ks-training}.
\begin{table}[H]
    \centering
    \caption{AR-DenseED and BAR-DenseED training parameters used for the Kuramoto-Sivashinsky system.}
    \label{tab:ks-training}
    \begin{tabular}{ll|ll}
    Training Parameters   &                                                  & SWAG~\cite{maddox2019simple} Parameters            &                                                  \\ \hline
    Optimizer             & ADAM~\cite{kingma2014adam} & Optimizer                  & ADAM~\cite{kingma2014adam} \\
    Weight Decay          & $0$                                                & Weight Decay               & $0$                                                \\
    Learning Rate         & $1e-3$                                             & Learning Rate              & $1e-10$                                            \\
    $\beta$ Learning Rate & $1e-3$                                             & $\beta$ Learning Rate      & $1e-6$                                             \\
    Exponential Decay Rate   &  $0.995$                       & Collection Rate     & $1$ epoch                                          \\
    Training Epochs       & $100$                                              & Models Collected           & $100$                                              \\
    Training Scenarios    & $2560$                                             & Deviation Matrix $H$ & 10                                               \\
    Mini-batch Size       & $256$                                              &                            &                                                 
    \end{tabular}
\end{table}

\subsection{Training Initial States}
\label{app:ks-initial}
\noindent
To train the model, we need to provide it a set of initial states or training scenarios to start at.
Simulator data could be used for this, however to stay consistent with the zero training data philosophy, we chose to use a truncated Fourier series with random coefficients.
We propose the use of:
\begin{gather*}
    u(x,0) = 2a\frac{w(x)-\min_{x}w(x)}{\max_{x}w(x)-\min_{x}w(x)}-a, \quad w(x) = \sum_{n=1}^3 \frac{\lambda_{n}}{n} \sin\left(\frac{n\pi x}{l}+c\right), \\
    \lambda_{n}=[1,\mathcal{N}(0,2),1], \quad c=2\pi\,\mathcal{U}[0,1], \quad
    a = \mathcal{N}(0,0.5) + a_{0},\\\quad l= L/(2k_{0}), \quad k_{0}=\floor{L/(2\pi\sqrt{2}) + 0.5},
\end{gather*}
where $a_{0}$ is the \textit{a priori} mean amplitude estimate (set to $2.5$), $L$ is the domain size and $k_{0}$ is the number of unstable modes which can be estimated given the domain length such that $k_{0}=L/(2\sqrt{2}\pi)$~\cite{cvitanovic2010state}.
This function is designed to provide a physically realizable initial condition for AR-DenseED to explore the physics of the K-S system.

% ======== Burgers' 1D Appendix ========
\section{1D Viscous Burgers' System}
\label{app:burger1d}
\begin{figure}[H]
    \centering
    \includegraphics[width=0.5\textwidth]{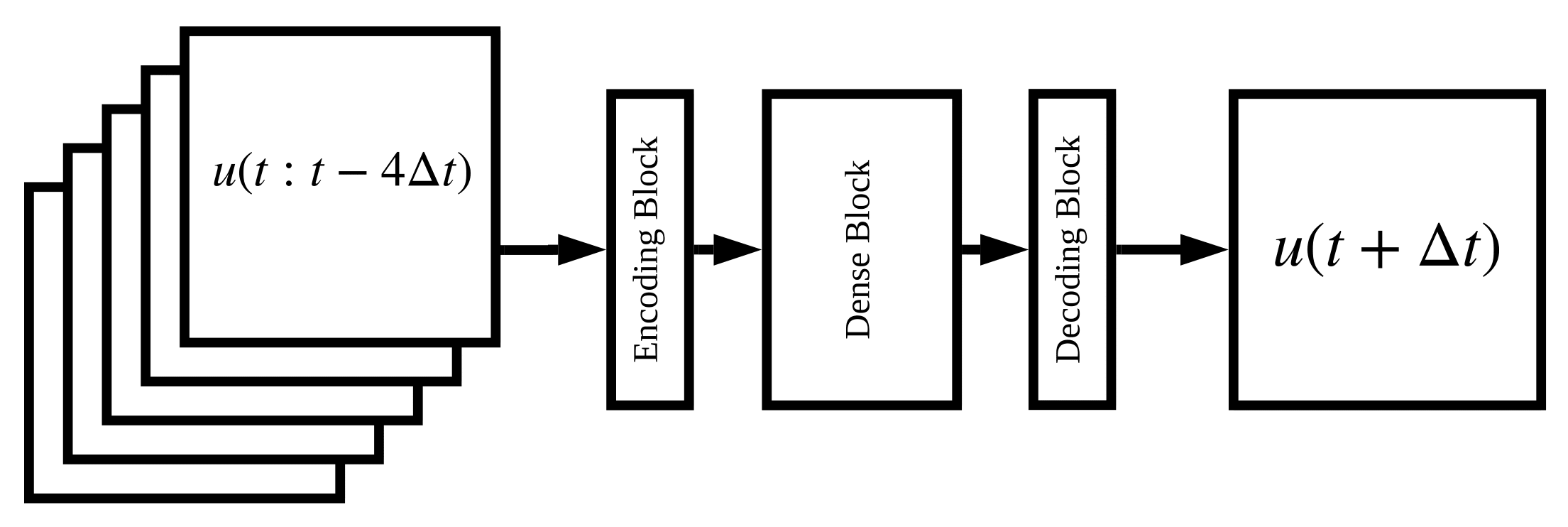}
    \caption{The AR-DenseED model with  $13442$ learnable parameters used for the 1D Burgers' equation.
    This model consists of an encoding convolutional block, single dense block with a growth rate of $4$ and a length of $1$ followed by a decoding block.
    The five previous time-steps are used as inputs.}
    \label{fig:app-burger1D-model}
\end{figure}

\noindent
The following appendix discusses details related to the model used to predict the 1D Burgers' equation in Section~\ref{sec:1dVisBurgers}.
Of the three systems present in this paper, the 1D viscous Burgers' system was found to be the most difficult to train.
This is most likely due to the sharp gradients in this system which carries over to the loss function resulting in exploding gradients during optimization.
The model for this system, depicted in Fig.~\ref{fig:app-burger1D-model}, is approximately three times as large as the model used for the K-S system.
Similar to the K-S model, the 1D-Burgers model encodes a given 1D input $\left\{\bm{u}^{n}, \bm{u}^{n-1}..., \bm{u}^{n-4}\right\}\in\mathbb{R}^{d}$ to a set of latent variables that are of dimensionality $\bm{z}_{i}\in\mathbb{R}^{d/2}$.
These latent variables are then decoded to the prediction $\bm{u}^{n+1}\in\mathbb{R}^{d}$.
More time-steps were used in the model input compared to the K-S system to increase learning stability.
Interestingly, if the model was too large, SWAG would fail to approximate a good posterior regardless of the learning rate resulting in sampled models being unstable during prediction.
Additional training parameters are listed in Table~\ref{tab:burger1D-training}.
\begin{table}[H]
    \centering
    \caption{AR-DenseED and BAR-DenseED training parameters used for the 1D Burgers' system.}
    \label{tab:burger1D-training}
    \begin{tabular}{ll|ll}
    Training Parameters   &                                                  & SWAG~\cite{maddox2019simple} Parameters            &                                                  \\ \hline
    Optimizer             & ADAM~\cite{kingma2014adam} & Optimizer                  & ADAM~\cite{kingma2014adam} \\
    Weight Decay          & $0$                                                & Weight Decay               & $0$                                                \\
    Learning Rate         & $1e-3$                                             & Learning Rate              & $4e-8$                                            \\
    $\beta$ Learning Rate & $1e-3$                                             & $\beta$ Learning Rate      & $5e-6$                                             \\
    Exponential Decay Rate   &  $0.99$                       & Collection Rate     & $1$ epoch                                          \\
    Training Epochs       & $100$                                              & Models Collected           & $90$                                              \\
    Training Scenarios    & $2560$                                            & Deviation Matrix $H$ & $30$                                               \\
    Mini-batch Size       & $256$                                              &                            &                                                 
    \end{tabular}
\end{table}

% ======== 2D Coupled Burgers' Appendix ========
\section{2D Coupled Burgers' System}
\label{app:burger2d}
\begin{figure}[H]
    \centering
    \includegraphics[width=0.5\textwidth]{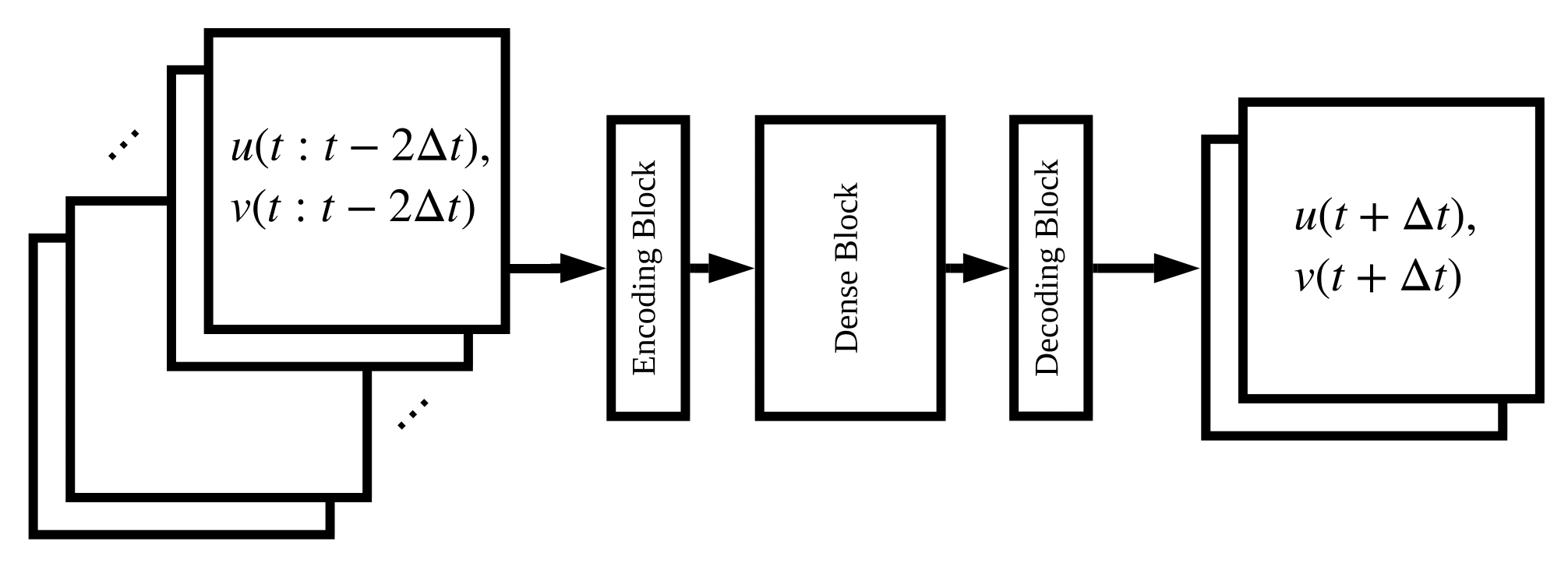}
    \caption{The AR-DenseED model with  $71953$ learnable parameters used for the 2D coupled Burgers' equation.
    This model consists of an encoding convolutional block, single dense block with a growth rate of $4$ and a length of $4$ followed by a decoding block.
    The five previous time-steps are used as inputs.}
    \label{fig:app-burger2D-model}
\end{figure}

\noindent
The following appendix discusses details related to the model used to predict the 2D coupled Burgers' equation in Section~\ref{sec:2dVisBurgers}.
As shown in Figure~\ref{fig:app-burger2D-model}, the model used for the 2D Burgers' system is very similar to the other two test cases.
The key difference is that the model now predicts two values: the $x$ and $y$ velocity components.
Similarly the model uses both velocity components as inputs. 
Thus when three previous time-steps are used in $\bm{\chi}^{n+1}$ the model has six input channels.
This model encodes a given 2D input $\left\{\bm{u}^{n}, \bm{v}^{n}, \bm{u}^{n-1}, \bm{v}^{n-1},\bm{u}^{n-2}, \bm{v}^{n-2}\right\}\in\mathbb{R}^{d \times d}$ to a set of latent variables that are of dimensionality $\bm{z}_{i}\in\mathbb{R}^{d/2 \times d/2}$.
These latent variables are then decoded to the prediction $\left\{\bm{u}^{n+1}, \bm{v}^{n+1}\right\}\in\mathbb{R}^{d \times d}$.
Although this system has some of the most complex dynamics, to our surprise we found that it was the easiest to train with significantly better stability and consistency than the other 1D systems.
We largely attribute this to the use of Sobel filters to approximate the gradients which help dampen higher frequencies preventing oscillations.
Additional training parameters are listed in Table~\ref{tab:burger2D-training}.
\begin{table}[H]
    \centering
    \caption{AR-DenseED and BAR-DenseED training parameters used for the 2D coupled Burgers' system.}
    \label{tab:burger2D-training}
    \begin{tabular}{ll|ll}
    Training Parameters   &                                                  & SWAG~\cite{maddox2019simple} Parameters            &                                                  \\ \hline
    Optimizer             & ADAM~\cite{kingma2014adam} & Optimizer                  & ADAM~\cite{kingma2014adam} \\
    Weight Decay          & $0$                                                & Weight Decay               & $0$                                                \\
    Learning Rate         & $1e-3$                                             & Learning Rate              & $3e-8$                                            \\
    $\beta$ Learning Rate & $1e-3$                                             & $\beta$ Learning Rate      & $3e-5$                                             \\
    Exponential Decay Rate   &  $0.995$                       & Collection Rate     & $1$ epoch                                          \\
    Training Epochs       & $100$                                              & Models Collected           & $90$                                              \\
    Training Scenarios    & $5120$                                             & Deviation Matrix $H$ & $30$                                               \\
    Mini-batch Size       & $128$                                              &                            &                                                 
    \end{tabular}
\end{table}

\clearpage
%\section*{References} % Remove this for arxiv

\bibliography{mybibfile}

\end{document}